# Using rxncon to develop rule based models

Jesper Romers*, Sebastian Thieme*, Ulrike Münzner and Marcus Krantz[1]

Institute of Biology, Humboldt-Universität zu Berlin, Berlin, Germany

* The authors contributed equally

[1]correspondence marcus.krantz@rxncon.org

**Abstract**

We present a protocol for building, validating and simulating models of signal transduction networks. These networks are challenging modelling targets due to the combinatorial complexity and sparse data, which have made it a major challenge even to formalise the current knowledge. To address this, the community has developed methods to model biomolecular reaction networks based on site dynamics. The strength of this approach is that reactions and states can be defined at variable resolution, which makes it possible to adapt the model resolution to the empirical data. This improves both scalability and accuracy, making it possible to formalise large models of signal transduction networks. Here, we present a method to build and validate large models of signal transduction networks. The workflow is based on rxncon, the reaction-contingency language. In a five-step process, we create a mechanistic network model, convert it into an executable Boolean model, use the Boolean model to evaluate and improve the network, and finally export the rxncon model into a rule based format. We provide an introduction to the rxncon language and an annotated, step-by-step protocol for the workflow. Finally, we create a small model of the insulin signalling pathway to illustrate the protocol, together with some of the challenges - and some of their solutions - in modelling signal transduction.

# 1. Introduction

Here, we present rxncon, the reaction-contingency language, as a tool to develop rule based models. Rule based modelling has been established as a powerful approach to modelling signal transduction networks [1, 2]. The main strength of the rule based approach is the adaptive resolution: Rules define reactions at an arbitrary resolution, where some features of the reactants may be specified while other are left undefined [3]. This is particularly useful when modelling the notoriously complex signal transduction networks [4, 5]. These networks typically transfer information by covalent modification or complexation of components. A single component may have many such state variables, which combines combinatorially into a large number of possible specific configurations, or microstates [6]. In contrast, empirical measurements are typically at the level of elemental, or macroscopic, states, which are only defined for a single state variable and hence correspond to non-disjunct sets of microstates [7]. By representing the network in terms of non-disjunct sets of reactions and states, the rule based approach avoids enumerating all the possible microstates in the system. This gives two powerful advantages: In comparison to microstate based models, the rule based models scale more efficiently and more accurately represent the underlying empirical knowledge [8].

The rxncon language takes this even further by fully adapting the network definition to empirical data. The network definition is separated between *elemental reactions*, i.e. the biochemical transitions that change elemental states, and *contingencies*, i.e. the dependencies of these transitions on elemental states (Fig 1) [9]. The elemental reactions and contingencies correspond more closely to empirical data than microstate or even rule based models (discussed in [8]). However, the rxncon language is closely related to the rule based languages: A rxncon network model fully defines a rule based model, which can be compiled automatically in the BioNetGen language ([10]; in preparation). In addition, rxncon offers a number of advantages compared to working directly with the rules: First, the rxncon language more closely relates to empirical knowledge, making it easier to write and maintain a rxncon model. Second, it is supported by an iterative model building – validation – gap finding & gap filling workflow that helps developing and debugging the network model [11, 12]. Third, it supports automatic model visualisation in compact graphical formats [13]. Fourth, the rxncon network constitutes an annotated knowledge-base that is human and machine readable, and hence easily reusable for further model construction, merging or analysis [8]. However, the rxncon network must be compiled into an executable model before it can be simulated.

In this chapter, we present a workflow to develop rule based models using rxncon. The core of the manual is a protocol covering five stages: (i) How to create a seed network that defines the model scope, (ii) how to turn the seed into a mechanistic rxncon network, (iii) how to convert the rxncon network into an executable Boolean model, (iv) how to use the Boolean model for qualitative network validation and debugging, and (v) how to export the final rxncon network into a rule based model.

This protocol is based on the second generation rxncon and its improved expressiveness and model generation semantics ([9, 12]; in preparation). We briefly introduce the language with syntax and semantics, and highlight where the network definition and model generation differ from the first version. Finally, we illustrate the language and the workflow by using rxncon 2.0 to develop a rule based model of a small part of the insulin signalling pathway.

**1.1 The rxncon language**

The rxncon language defines biochemical reactions in terms of components, elemental states, elemental reactions and contingencies (Fig 1; [9]). The *components* are the independent agents in the network: The proteins, the second messengers, etc. They correspond to the molecules in rule based models. The *elemental states* are fully defined and indivisible state properties of one or more components, such as a specific covalent modification on a specific site, an interaction between two specified domains of two proteins, or the absence of any modification or interaction on a specific residue or domain, respectively. In terms of rule based models, an elemental state corresponds to the state at one specific site. The elemental states are the actual information carriers, as they define the mechanistic changes in the signalling network that transfer information. The *elemental reactions* are indivisible reaction events that produce or consume one or more elemental states (Table 1; Fig 2). Elemental reactions are defined as fully decontextualized *skeleton rules* [9], meaning that only the catalyst (if any) and the reaction centre are defined [14]. In other words, only the core components and the elemental states that change in the reaction are defined. Finally, the *contingencies* define the prerequisite states that are required for the reaction. In terms of rule based models, the contingencies correspond to the reaction context – i.e. the elemental states that are required for, but do not change through, the reaction. In rxncon, both elemental reactions and contingencies are defined in terms of elemental states. The contingency definition can be made arbitrarily complex by using Boolean combinations (AND, OR NOT) of elemental states, which can be used to define structured complexes when necessary. Hence, rxncon can be used to build a model that is as complex as necessary (as defined by empirical data), but not more complex.

There are two strategies to simulate rxncon networks. First, the bipartite network definition can be compiled into a bipartite Boolean model that can be simulated without further parametrisation [12, 15, 16]. In this case, elemental reactions and elemental states appear as distinct sets of nodes with two sets of update rules: The state node updates are determined by the elemental reactions, and the reaction nodes by contingencies. Based on a set of standard assumptions, each rxncon network defines a unique Boolean model with fully specified update rules. Below, we will use this to validate the network structure before generating a rule based model. Second, the reaction and contingency information can be compiled into a rule based model ([10]; in preparation). The elemental reactions provide the skeleton rules that define the reaction centre. The skeleton rules are refined by applying contingencies, which define the reaction context. The complete model creation process is implemented in the rxncon compiler.

These two modelling strategies differ in detail and perspective: The rule base model creates reaction rules at the level of molecules, while the Boolean simulates the reaction network at the level of the system. Consequently, states that are mutually exclusive at the level of individual molecules may coexist in the Boolean model. In addition, the rule based model requires rate law assignments and parametrisation, in contrast to the bipartite Boolean model that can be simulated without quantitative information. Hence, the second can be used to validate and debug the network structure before the generation and parametrisation of a rule based model.

This protocol below is based on rxncon 2.0, which is described in detail elsewhere ([9, 12]). We have updated the rxncon language to improve the expressiveness and model generation semantics. In particular, the rxncon now supports definition of structured complexes that can distinguish e.g. cis- and trans-phosphorylation across homodimers, which was not possible in rxncon 1.0 [10]. Furthermore, the reaction definition syntax has been refined and now explicitly includes neutral elemental states, such as unmodified residues and unbound binding domains (Fig 2; Table 1). Together with a flexible definition of elemental reactions through skeleton rules, these changes make it possible to define essentially any reaction rule at the level of biomolecular site dynamics using the rxncon language.

However, there are certain limitations to rxncon. First, the rxncon network defines a qualitative network model (QlM;[8]). The network can indeed be converted into a ready-to-run rule based model, but this model has trivial parameters and initial conditions [12]. Quantitative information must be added directly in the rule based model code. Second, the rule based modelling languages can encode processes that are not possible to express in elemental reactions and contingencies, e.g. via functions in the BioNetGen language [3]. Finally, the rxncon language is reaction focused at the molecular level. It is difficult or impossible to meaningfully model higher level mechanisms, such as e.g. actin polymerisation or vesicle transport. Nevertheless, within the scope of biomolecular site dynamics, rxncon provides a powerful approach to model development.

### 1.2 Developing a rule based model with rxncon

Here, we present a detailed workflow for developing rule based models with rxncon. The workflow is inspired by the analogous workflow for metabolic network reconstruction [17], but uses methods that are tailored to the very different properties of signal transduction networks ([11]; reviewed in [8]).

The workflow can be broken down into five steps: (i) Scope definition and network seed creation, (ii) refinement of the seed into a mechanistic QlM, (iii) conversion of the QlM into an executable bipartite Boolean model (bBM), (iv) evaluation of the bBM and hence the functionality of the QlM, and (v) conversion of the QlM into a rule based model. Steps (ii) – (iv) are typically performed iteratively, until the bBM reproduces the expected *in vivo* functions *in silico*. The objective of the workflow is to create a well annotated knowledge base, that qualitatively reproduces the system level function, and to convert this knowledge base into a rule based model. For clarity, we present this as a sequential workflow. However, this would in practice be an iterative process where the literature is used both to refine the scope and the mechanistic model simultaneously.

The process starts with a biological question or topic of interest. The process is easier when this is the function of one or more pathways, as these have defined functions in terms of inputs (what does the process respond to or require?) and outputs (what does the process do?). Knowledge of inputs and outputs will help defining the scope, interpret the literature, and to validate the final model. In this case, we know which macroscopic input-output behaviour to expect, and can use this knowledge to evaluate the completeness and accuracy of the microscopic biochemical reaction network we defined. The rxncon language is also suitable for true bottom up model building: Elemental reactions are independent of other elemental reactions, and contingencies are independent of other contingencies. Hence, the language is highly composable, which greatly facilitates network reconstruction from fragmented information. However, the current strategy for validation relies on comparison to known macroscopic input/output behaviours. Therefore, we will assume that the model is built in the context of a pathway or process with defined inputs and outputs.

Once the scope is defined in terms of inputs and outputs, it is helpful to collect information on which components are needed to convey the information from the input to the output. This information is typically available from genetic data, where studies have shown that certain signals or processes require certain components. These studies can also be used to derive epistatic information, i.e. in which order components act in the pathway, and this can be used to compose a draft network (Fig 1). The scope in terms of inputs/outputs and components can typically be collected from review papers. In practice, this is the reading up step of getting in to the topic, and the formal creation of a network seed is optional.

The second step is to create the mechanistic model. To turn the network model into a mechanistic QlM, two layers of information are required (Fig 1): First, what are the actual biochemical transitions in the network? Second, what are the causal relationships between these transitions? Practically, the easiest is to start with pairs of components that are thought to be directly connected to each other and search for biochemical reactions that connect these components. To define the elemental reactions and states, three parts of information are needed; the components involved, the type of reaction between them, and the state the reactions result in. Typically, the state is observed and the reaction type needs to be inferred. To isolate indirect and direct effects, reactions should typically be monitored *in vitro*. The most reliable data for building the reaction layer is *in vitro* biochemical data.

The causal layer is more elusive. The contingencies are defined as the effect of one or more elemental states on an elemental reaction. To infer causal effects of previous state modification, e.g. the role of phosphorylation in activation of a protein, more complex data is needed. Ideally, this information is based on a combination of genetic and biochemical *in vitro* experiments, where mutant forms of proteins are isolated and analysed. For example, candidate phosphorylation residues can be substituted to mimic or prevent phosphorylation, and the activity of the mutant protein compared to the wild type protein. While the states needed to infer the reaction layer can be measured in high-throughput (although with the risk of scoring indirect effects), the information required for the contingency layer can to date only be generated with dedicated low throughput experiments.

We recommend using primary literature to build the QlM. Review articles are excellent to define the scope and to create a network seed. They can also be used to build a mechanistic model, as we do in the example of the insulin pathway. However, the objective is a detailed mechanistic model, based on highly specific statements. This level of detail is often missing in review papers, or in the text in original papers, and we find it invaluable to return to the actual empirical data. In our experience, high quality curation often involves reinterpretation of the underlying experimental evidence.

The output of the second step is a QlM in the rxncon language. It is defined as a table of elemental reactions, which defines the reaction layer, and a table of contingencies, which defines the causal layer together with the reactions. The model can be visually inspected at both levels of detail. The reaction layer can be visualised as a rxncon reaction graph (Fig 1B; Fig 3B). Each edge corresponds to a reaction, and a chain of reactions is required, but not sufficient, to allow information transfer. The full QlM can be visualised in the rxncon regulatory graph. This bipartite graph visualises elemental reactions and states as nodes connected by reaction and contingency edges. Paths from inputs to outputs are required, but again not sufficient, for information transfer (rRG; Fig 1D; Fig 4). It is only meaningful to proceed to model generation and evaluation if there are paths from the inputs to the outputs in the rRG.

The third step is the model generation. In this step, the rxncon QlM is converted into a bBM based on the rRG network. This bipartite network has two node types; elemental reactions and elemental states, which follow two distinct update rules, as described in detail elsewhere [12]. The bBM is fully defined by the QlM, given a set of standard assumptions, and the model generation can be performed automatically.

The fourth step is the model evaluation. In this step, the bBM is used to simulate the network in response to changing inputs, and the simulation results compared to the known *in vivo* behaviour. For each input/output pair, the process tests if the output is responsive to the input, and if the signal is transmitted by the pathway. If not, this discrepancy is used for targeted model improvement, and the process returns to step two or even one to extend or refine the QlM. If all tests of the bBM pass, the QlM is considered qualitatively functional and ready for conversion into a rule based model.

The fifth step is to convert the QlM into a rule based model. The rxncon framework code can compile a set of rxncon statements into a rule-based model formulated in BNGL, as will be described in detail elsewhere (in preparation). Here, we will briefly touch upon several elements that constitute a rule-based model and their relation to a rxncon model.

To determine the Molecule Types, the elemental states appearing in the rxncon model are grouped by component. The modification states are then grouped by residue: these determine the molecule's sites and their possible modifications. The interaction states' domains are also mapped to molecular sites.

The model is seeded by default with "neutral molecules": this means that all modifications are in the neutral state and all molecules are unbound.

Each elemental reaction, combined with its contingencies, is then translated into one or more rules. The reaction's skeleton rule provides the rule's centre, whereas every solution to its contingency's Boolean equation provides a possible context for that rule. For example, a straightforward elemental state requirement will provide only one such context. On the other hand, a Boolean OR statement of three phosphorylation sites will lead to seven contexts due to combinatorics. To translate Boolean NOT statements, the complementary values at a molecular site are taken, connected via an OR.

Finally, the output states in rxncon are mapped on BNGL observables.

## 2. Material

*1. The rxncon framework*

The rxncon framework is distributed in two varieties: as a full framework, including a web-based front-end, and as a Python library for inclusion in an existing pipeline. Both varieties require a preinstalled Python (v3.5 or higher)[1]. The rxncon software is developed open source, and distributed under the lGPL license. The code is available from Github (https://github.com/rxncon/rxncon).

Both the standalone Python 3 library and the graphical user interface can be installed from the Python Package Index (PyPI) via the pip tool[2]. All necessary dependencies are automatically resolved and installed.

*2. The rxncon input file: spreadsheet template*

The rxncon models are most easily defined in SBtab compatible spreadsheet files. A template can be downloaded from GitHub (template.xls; https://github.com/rxncon/models/). To work with this file, we recommend MS Excel or Gnumeric.

To build a rxncon network, the user needs to fill two lists: The reaction and contingency lists, as described in detail in section 3 below. The reaction definition is supported by two additional sheets: the ReactionTypeDefinition, which lists all valid reaction types, and the ModificationTypeDefinition, which lists all valid modification types. Both these lists are extendable, enabling easy extension and customisation of the rxncon langague.

In the reaction list, the elemental reactions are defined as two specs and a reaction type (see: [9] for details). Each spec corresponds to a component and possibly a domain and/or residue definition. Each Spec definition is separated into up to three columns, as exemplified by component A :

!ComponentA:Name: Name of reaction partner A (the subject), e.g. a protein.

!ComponentA:Domain (optional): Domain of reaction partner A.

!ComponentA:Residue (optional): Residue of reaction partner A.

The "!Reaction" column specifies the reaction type, which must refer to a unique reaction key in the ReactionTypeDefinition sheet (in column "!UID:ReactionKey"; see below).

In addition, the network can be annotated through the following columns. Entries in those fields does not influence parsing or processing, but can be used to increase the quality and reusability of the model:

!Quality: Quality of the empirical evidence of the reaction, e.g. the type of experiments or your confidence in the reaction assignment.

!Literature:Identifiers:pubmed: Identification tag of source of information. We prefer using PubMed identifiers, but any unique identifiers can be used.

!Comment: Comment the reaction if necessary.
*Note: It is advisable to clearly distinguish any hypothetical reactions added to debug the model from high confidence reactions based on empirical data to make sure the actual knowledge-base can be separated from pragmatic model improvements.*

The "!UID:Reaction" column is a concatenation of the information from the Component & Reaction columns into a unique reaction ID. It will be filled automatically by excel. Make sure to copy the functions from the rows above, but do not edit them.

In the contingency list, the contingencies are defined in an object-verb-agent passive clause over the following columns:

!Target: An elemental reaction or output that is regulated. The target column is also used to define Boolean states (see Fig 2). Outputs are defined as text strings within hard brackets, e.g. "[Output]", Boolean states are defined as text strings within pointy brackets "<BooleanState>", and elemental reactions are defined by the unique ID from the reaction list (in column "!UID:Reaction").
*Tip: In spread sheet editors like Excel, it is convenient to link the cell to the reaction in the reaction sheet. This way, the contingency will be updated as soon as the reaction is updated. This is not necessary but makes the editing more convenient.*

!Contingency: The contingency required for the reaction in "!Target". There are six valid contingency types: "!" Absolute requirement; "x" Absolute inhibition; "K+" Positive influence (increase of reaction rate); "K-" Negative influence (decrease of reaction rate); "0" No effect; "?" No known effect. Boolean definitions take Boolean operators instead; "AND", "OR" or "NOT".

!Modifier: The elemental state, Boolean state or input that the "!Target" depends on. Inputs are defined as text strings within hard brackets, e.g. "[Input]", Boolean states are defined as text strings within pointy brackets "<BooleanState>", and elemental states are defined by the elemental state string.[3, 4]

As in the reaction list, there are three columns used for model annotation: "!Quality", "!Reference:Identifiers:pubmed", and "!Comment".

The ReactionTypeDefinition sheet holds all valid reaction types. Each reaction type needs a unique ID[5] and is defined by a skeleton rule, a type and resolution definition of the components, and a type of directionality.

- "!UID:Reaction ": Free text name of reaction.
- "!UID:ReactionKey": Unique identifier for the reaction type.
- "!BidirectionalVerb": Typically "no", if "yes", reactions are generated in both directions.
- "!MolType": The type of molecule the reaction targets, e.g. Protein, mRNA, Gene or Any.
- "!Resolution": The locus resolution needed: Component, Domain or Residue.
- "!SkeletonRule": Semantic definition of the reaction type.

To create new reaction types, it suffices to add a new entry to the list as long as the modification type is declared in the ModificationTypeDefinition sheet. If not, the new modification type must be added to this list: Make sure the ID is unique and that the type and resolution definitions are consistent with the skeleton rule. For more details, see [9].

*3. Model visualisation: Cytoscape*

The visualisation of rxncon networks will use Cytoscape, which can be downloaded from cytoscape.org. The graphical styles used for the reaction and regulatory graphs can be downloaded from GitHub (rxncon2cytoscape.xml; https://github.com/rxncon/tools).

*4. Bipartite Boolean simulation: BoolNet & R*

The logical simulation of rxncon networks uses BoolNet, an R package.[6] To use these tools:

i. (Optional) Download and install R-studio (https://www.rstudio.com).[7]
ii. Make sure you have R installed.[8]
iii. Install the BoolNet package.[9]

We have prepared an R-script for the simulation ("BoolNetSim.R") which can be downloaded from: https://github.com/rxncon/tools.

*5. Agent based simulation: NFSim & BioNetGen*

The rule-based models generated by the rxncon framework require either BioNetGen or NFSim to simulate. We recommend NFSim, which contains BioNetGen but also supports agent based simulation, which will be necessary for larger networks. NFSim can be downloaded from http://michaelsneddon.net/nfsim/download/.[10]

## 3. Method

### I. Define the model scope and create a network seed.[11]

1. Define the activator (input) of the pathway and the expected behaviour (the output).[12]
2. Define the components that connect the input(s) with the output(s). This information is typically available from review papers or paper introductions.
3. Define the sequential order in which the components act, as far as possible. This is a refinement of (2), which helps to narrow down the search for mechanistic connections. If possible, determine which components are directly connected.
4. Optional: Create a conceptual network in which you sketch important events of your pathway. These events can correspond to phenomenological observations and thus do not need to be on a molecular level. A conceptual network might help alongside to keep an overview.

**Exempli Gratia: The insulin signalling pathway – part 1.** Our aim is to build a rule based model of the initial events in the insulin signalling pathway. Naturally, we chose insulin as the input. Choosing the outputs is less obvious, but we decide that we are interested in the initial signalling events. We hence chose the activation of the phosphoinositide 3-kinase (PI3K) and of Ras. We base these choices (and, for simplicity, all further work in this example[13]) on a single review paper [20]. Now, we identify the important components that connect the input with the output: The insulin receptor (IR) in its homodimeric form (the insulin receptor can also heterodimerise, but we leave this out); the insulin receptor substrate (IRS; again multiple forms exist), the Shc protein; the growth factor receptor-bound protein 2 (Grb2); the son of sevenless homolog (SOS); and the PI3K. We added phospholipids to model plasma membrane binding of some of the components. Finally, we sort the components in order and include the direct connections we gathered from the review. The scope and seed are summarised in figure 3A.

### II. Creation of a mechanistic model

The mechanistic model is built in two layers: The reaction list defines the possible events in the network as decontextualised reactions between pairs of components (or intramolecular events in a single component), while the contingency list defines the contextual constraints on these reactions. The example model built here as well as a template is available for download from https://github.com/rxncon/models/.

*Definition of the reaction layer*

The reaction layer specifies which elemental states are produced or consumed by the interacting components. It is built in the reaction list of the spreadsheet template (see material section above).

1. For each presumed component pair, search the literature for reactions that connect the pairs.[14]
2. Determine which reaction type best describes the event [15] (see the ReactionTypeDefinition sheet). If no reaction type matches the event, define a new reaction type.[16]
3. Enter the reaction type in the column "!Reaction". [17]
4. Specify the component(s) taking part in the reaction. [18, 19]

    - The name of the components should be specified in columns "!ComponentA:Name " and "!ComponentB:Name".[20]

    - (Optional) The name of the interaction domains should be specified in "!ComponentA:Domain" and "!ComponentB:Domain". [21, 22]

    - (Optional) The name of the target residues should be specified in "!ComponentA:Residue" and "!ComponentB:Residue". [23]

5. Annotate the reaction entry

    - Optional: Use the column "!Quality" to specify the type of experiment and/or your confidence in the evidence.

    - Optional: Use the column "!Literature:Identifiers:pubmed" to annotate the source.[24]

    - Optional: Add comment in "!Comments" if desired.

6. Repeat these steps until all the reactions you found are considered in the reaction list. It is often necessary to extend the search beyond the initial scope.

The rxncon model is processed by the rxncon compiler, which can be accessed either from the command line or through a graphical user interface (in preparation). To use these rxncon compiler through the graphical user interface:

7. Create a new project:

    „New"-Button on left sidebar and choose source:

    a. „ spreadsheet file"
        i. Set the project name.
        ii. Click on 'choose file'.
        iii. (Optional) Add a comment for this project in the project comment box.
        iv. Press „Upload File".[25]

    b. „ Text input"
        i. Set the project name.
        ii. Enter Quick input.

iii. (Optional) Provide a comment.
iv. Press „Save".
8. Update a project (spreadsheet input only):
   a. Click on the plus symbol of the respective project in the project sidebar.
   b. Choose your file, enter a comment if applicable, and click update.
9. Load a project (activate it):
   a. Load the desired network to the workspace by clicking 'load file' (the bolt symbol).
   b. Load an older version of a project by first opening the detail view (the eye symbol), then push the load button of the respective version.

*Check for gaps in the reaction layer*

Next, we visualise the reaction layer with the rxncon reaction graph. This graph visualises all components, with domains and residues, as nodes and all reactions as edges. Information can only pass from one component to another if there is a path of reactions connecting the two.

To visualise the reaction topology of the network:

10. Create the reaction graph either from the command line or through the graphical user interface.
    a. To create the graph from the command line:
       Call the "rxncon2reactiongraph" script with the excel file as argument.[26]
    b. To create the graph through the GUI:
       i. Load the project.
       ii. In the top navigation bar, choose „Visualisation" → „Reaction Graph".
       iii. (Optional) Provide a comment.
       iv. (Optional) Reuse previous layout.
           To do this, import an .xgmml file with layout information. [27]
       v. Create the graph by clicking "Create Graph".
       vi. Retrieve the file by clicking the "Show" button in the respective detail view (the eye symbol).
10. Load the .xgmml file in Cytoscape (>v3.4.0)
11. Import the visual style file for rxncon.[28]
12. Set the visual style to rxncon reaction graph.
13. Move the nodes to retrieve an appealing network layout.

*Note: The domains and residues are intended to be laid out adjacent to the component node. This does not work well with automatic layouts.*[29]

14. Inspect the connectivity between the components.
    The objective is to determine if there is a mechanistic (reaction) path from each input to each output that responds to that input. To determine this, look for connections from the most upstream to the most downstream component in each information path (input-output path). The inputs and outputs will not be connected yet, as these connections are defined as contingencies (see below; Fig 1B).

15. If the network is not connected, extend the network with further reactions by either gathering additional empirical information or stating hypothetical reactions until you obtain a connected graph.

16. Only proceed if there is a path from the most upstream to the most downstream components. [30]

**Exempli Gratia: The insulin signalling pathway – part 2.** Using the review [20], we identify the elemental reactions that connect the components with each other. We are searching for references to direct mechanistic connections, like "Interactions of insulin with IR have been studied in greatest detail …", which we interpret as a direct binding of insulin to the insulin receptor (IR). In the reaction list of the rxncon sheet, we put IR as "ComponentA[Name]", the reaction type "i" for interaction and insulin as "ComponentB[Name]". To make the reactions and states elemental, we want to assign a domain name to each of the component. As we found no information on where insulin binds the IR, we called the domain on IR "lig" (for ligand) and entered it as "ComponentA[Domain]". Next, we find the information: "Shc proteins are … substrates of the IR and IGFR…" which identifies the phosphorylation of Shc by IR. We enter "IR" as "ComponentA[Name]", the abbreviation "P+" for phosphorylation as the reaction type and "Shc" as "ComponentB[Name]". In this case, the review mentions two distinct target residues the amino acid Y317 and YY239 (Y239, Y240 or both, which is not clear from the text; we refer to this site as YY239) that are phosphorylated: "All three Shc isoforms are tyrosine phosphorylated … on two distinct sites (YY239/240 and Y317)." Hence, we need two distinct reactions that differ only in their target residues. In the case of IRS, the review is less specific and refers only to tyrosine phosphorylation. Without further information, we enter this as a single site but suspect the review refers to multiple such sites that may have overlapping functions. We also find out that these residues form the core of binding domains, so we assign domain names (bd, bd1 or bd2 for "binding-domain", as the review doesn't specify domain names). We proceed through the review to build the reaction list presented in table 2. Once we consider the reaction list complete, we proceed to visualise it. To this end, we create a reaction graph of the network and check if we connected insulin (our most upstream component, as there is no input outside the model in this case) to the most downstream components; PI3K and SOS. The graph is connected (Fig 3B), and we proceed to the causal layer.

*Contingency definition*

The contingencies specify which (combinations of) elemental states have a regulatory effect on which reactions.

17. Use primary literature to find contextual information.[31] For each elemental reaction, define which (combination of) elemental state(s) of components A and B the reaction depends on.
    a. In the literature, look for data that identifies the active form of the components[32].
    b. Define the active form in terms of an elemental state or a Boolean combination of several elemental states, e.g. modification(s) and bonds to other components.[33]
    c. Determine which part of the requirement is already defined in the reaction (e.g. components, source states). This should **not** be defined as contingencies.
18. Define the contingencies in the "contingency list" in the rxncon template:
    a. Add the reaction that requires a contingency to the column "target".
    b. Define the type of contingency in column "Contingency".[34]
    c. In "Modifier", place the elemental state, Boolean state or input that regulates the reaction. Each reaction can use as many contingencies as needed. For combinations of contingencies, Boolean states (identified by pointy brackets < >) can be used to code complex requirements.[35] The inputs (identified by hard brackets [ ]) define the regulation of the system by external factors. [36]
    d. Add structure indexes if necessary. Under some circumstances, component names are not sufficient to define contingencies. In particular, this is the case when the component name is not unique, i.e. in case of complexes containing more than one subunit of the same kind.[37] In these cases, a position label must be added to the contingencies (Figure 2).
19. When Boolean states are used as modifiers, the Boolean state must be defined in the contingency list. To do this:

a. Enter the Boolean state name (marked by pointy brackets: < >) in the target column.
b. Enter the Boolean operator in the contingency column. Valid operators are AND, OR and NOT.
c. Enter either an elemental state, another Boolean state, or an input as modifier.
d. In structured complexes, define equivalences when necessary. [38]
20. Define the contingencies for the output of the network. In the context of rule definitions, the outputs will define the observables.
   a. Enter the output name (marked by hard brackets [ ]) in the target column.
   b. Enter the contingency as for elemental reactions above.
   c. Enter the modifiers as for elemental reactions above.

*Visualise the regulatory structure and check for gaps*

21. Create a regulatory graph. The process is analogous to the creation of the reaction graph (see (7) above). [39, 40]
22. Visualise the graph in Cytoscape. [41]
23. Inspect the graph:
    Determine if there is a directed path from each input to each output.
    If the network is not connected, extend the network with further reactions and/or contingencies by either gathering additional empirical information or stating hypothetical reactions until you obtain a connected graph and repeat the steps in this section.
24. Only proceed to model generation if there is a path from each input to each output.

**Exempli Gratia: The insulin signalling pathway – part 3.** To add the causal layer, we return to the literature (review) and search for information on how reactions are regulated. For example, we learn that Grb2 binds Shc when Shc is phosphorylated on residue(s) YY239. The SH2 domain in Grb2 binds the phosphotyrosine at position 239/240 in Shc. We have already defined the reaction that Grb2, via its SH2 domain, binds Shc at the bd2 domain, which contains the YY239 residue. However, this residue must be phosphorylated. Consequently, we specify a contingency of the protein-protein interaction on the phosphorylation: the contingency sheet, we put the ppi-reactions between Grb2 and Shc in the "Target" column. To specify phosphorylation of this residue as a requirement, we put the state "Shc_[bd2(YY239)]-{P}" in the "Modifier" column and enter absolute requirement ("!") in the "Contingency" column. Based on the review paper, we cannot identify further contextual constraints on these bonds, though further requirements might appear if more literature is considered or new empirical data becomes available. Insulin binding is more difficult to define. Insulin binding requires receptor dimerisation, but only one insulin molecule can bind the receptor. In rxncon, we cannot define an interaction with a complex, so the reaction is "IR_[lig]_i_insulin" (IR interacts with insulin). Adding the requirement for dimerisation would be easy (IR_[lig]_i_insulin ! IR--IR), but we need a more complex statement to account for the fact that the dimer only binds a single insulin molecule. In the end, we settle for a Boolean state: <IR-empty> = *IR@0--IR@1 AND IR@1_[lig]--0* (see Figure 2). Here we make use of structured complexes, where the identity of each subunit is labelled by an "@" followed by a unique number. The reactants have index 0 (IR) and 1 (insulin), and we define in the contingency that the IR in the reaction is equivalent to the IR at position 0 in the Boolean state "<IR-empty>#IR@0=IR@0" (each Boolean has its own name space, see [9] for more details). Finally, as an example of an output contingency, we define the requirements for Ras activation. It is one of two possible outputs of the network. From the review, we gather that at least one of two conditions must be fulfilled: 1) Grb2 must be bound to IRS and SOS; 2) Grb2 must be bound to SOS and Shc. In this situation, it is convenient to use nested Boolean expressions to describe these conditions in the contingency list. First, we define with an OR statement that either complex 1) or 2) is needed. Next, we define the two Grb2 complexes. Based on the review, we continue to define contingencies until we consider the list complete (table 3; the complete model can be found in Sup File 1). At this stage, we visualise the reaction-contingency network in the regulatory graph (Fig 4). The objective is to determine if every input is connected to every output. In this model, we did not define an external input, instead using the neutral state of insulin (the binding reaction between IR and insulin is our most upstream node in the regulatory graph). Consequently, we check if there is a directed path from this node to each of the two outputs. This is the case and we proceed to the next step.

**III. Bipartite Boolean Model (bBM) generation and simulation**

The next step is to create an executable model from the QlM by exporting the rxncon network to a bipartite Boolean Model (bBM). The model generation uses an algorithm described in detail elsewhere [12].[42]

1. Create the bipartite Boolean Model file. This can be done via the GUI or from the command line.

a. To generate the model from the command line:

Call the "rxncon2boolnet" script. [43]

*Note: The model can be generated with different options.* [44]

b. To generate the model from the GUI:
   i. Load the respective project.
   ii. In the top navigation bar, choose "Model Export" → "Boolean Model".
   iii. (Optional) Provide a comment.
   iv. (Optional) Change default export parameters.
      *Note: The model can be generated with different options.* [45]
   v. Create the BoolNet files by clicking "Create model files".
   vi. Retrieve the files by clicking the "Show" buttons in the respective detail view (the eye symbol).

The model creation generates three different files:

- [model].boolnet:         The model file, encoded as symbolic update rules
- [model]_symbols.csv:     The key to the symbols in the model file
- [model]_initial_vals.csv The initial state of the network

where "[model]" is the file name (without extension) of your rxncon model.

2. Inspect and/or adjust initial conditions. [46]
3. Simulate the file in BoolNet using RStudio: [47]
   a. Save the network files and the R script into a single directory.
   b. Start RStudio.
   c. Open a new project and create it in the directory where you saved your files.
   d. Make sure your model files are located in the project folder. [48]
   e. Open the R script. Set the filePrefix in the R script to [model]. [49]
   f. Execute the entire script by selecting all text (ctrl+a) and pressing ctrl+enter.
   
   The script generates five files:
   
   1. [model].pdf                     The simulation trajectory, graphical
   2. [model]_trajectory_first.csv    The simulation trajectory, values
   3. [model]_2.pdf                   Second simulation trajectory[50], graphical
   4. [model]_trajectory_second.csv   Second simulation trajectory, values
   5. [model]_new_attractor.csv       The attractor reached[51]
   
   where "[model]" is the file name (without extension) of your rxncon model.

4. Inspect the simulation results: [model].pdf; [model]_2.pdf. The first file displays the path to the attractor, the second the type of attractor (point = 2 columns; cyclic > 2 columns).

5. Save the file under new names before rerunning the script. The files will be overwritten.
6. To repeat the simulation from the new steady state, rename the "[model]_new_attractor.csv" file "[model]_initial_vals.csv.
7. (Optional) Adapt the input states to perturb the model.
8. Repeat from (2) above.

**Exempli Gratia: The insulin signalling pathway – part 4.** We export to a bBM using the rxncon framework (Sup. File 2-4). The model generation creates three files containing the model, the initial values, and a key between model IDs and the reaction and state names. The initial values file is already set to the default initial conditions: All neutral states are true (1). All other states, all reactions and all inputs are false (0). As we would like to use insulin as an input, we also set the neutral state of insulin to false (0). This default starting point allows the system to reach steady state in the absence of the signal, and we can then see how it reacts to the perturbation. We simulate the network to steady state and save the steady state, and as expected the output signals [PI3K] and [RAS] remains off. In the second round of the simulation process we set the neutral state of insulin (insulin_[IR]--0) to true (1). At the end of this simulation, the output signals [PI3K] and [RAS] are true. In round three, we turn insulin off again and expected to get the same attractor as in round one, but the outputs remain on. As we do not reach a steady state we have seen before, we proceed to a fourth round of simulation. We change insulin back to True and, as expected, the outputs are still on. As both the inputs and outputs are the same in steady state 2 and 4 (this was not the case for any previous pair of steady states), we proceed to compare the attractors. They are identical, and we proceed to the evaluation step, eager to figure out why the network cannot turn off after the initial activation.

**IV. Model validation and improvement**

The aim of the model validation is to identify which input-output pairs are connected by a functional information transfer path. To this end, we will turn inputs on and off and examine which outputs respond to these changes.

1. Create the bipartite Boolean model and run the first simulation without editing the "[model]_intitial_vals.csv" file.
2. Remove or rename the old initial_vals file, and replace it with the "[model]_new_attractor.csv" file by renaming it "[model]_initial_vals.csv". Turn the input of interest on by setting the value to True (= 1). [52]

3. Move or rename previous output files to save them from being overwritten.
4. Rerun the simulation and determine which output changes.
5. Remove or rename the old initial_vals file, and replace it with the "[model]_new_attractor.csv" file by renaming it "[model]_initial_vals.csv". Turn the input of interest off again by setting the value to False (= 0).
6. Move or rename previous output files to save them from being overwritten.
7. Rerun the simulation and determine which output changes. Compare the steady state to that after the first simulation (1). If the steady states are the same[53], we are done with this particular input. If not, continue.
8. Remove or rename the old initial_vals file, and replace it with the "[model]_new_attractor.csv" file by renaming it "[model]_initial_vals.csv". Turn the input of interest on by setting the value to True (= 1).
9. Move or rename previous output files to save them from being overwritten.
10. Rerun the simulation and determine which output changes. Compare the steady state to that after the second simulation (4). If the steady states are the same, we are done with this particular input. If not, continue simulating with iterative input states until the attractor reached has been seen before.
11. Evaluate the response of the effect of the input across all outputs and simulations, to determine which outputs are regulated by this input (according to the model). [54]
12. Compare to the known macroscopic input-output behaviour of the network. Is it reproduced by the network model?
    a. If yes: the model is considered functional – proceed to generate the rule based model.
    b. If not: use the simulation results to identify where the signal is blocked. If often helps to work from both directions: [55]
        - How far does the signal reach from the input?
        - Which reactions/states should have responded in order to affect the output?
13. Use the insight from the model analysis to improve the mechanistic model (Phase II). To debug the model:
    a. Search for the most upstream reaction or state node that does not respond. For most models, this is most easily done by sorting the heatmap on transitions and inspecting it visually (see Fig 5).
    b. If the most upstream node is a state, a reaction is either missing or off. [56]
    c. Check if a valid reaction is included in the network, but blocked by contingencies. If so, re-examine these contingencies.

d.  If not, add a reaction targeting that state. Use this candidate reaction for a targeted literature search. If nothing can be found, consider experimental validation or enter it as a hypothetical reaction that needs to be verified.

e.  If the most upstream node is a reaction, there is a problem with the contingencies. Re-examine the contingencies to see how to make the reaction responsive in both directions – use the new contingencies for targeted literature search, experimental validation or to enter hypothetical mechanisms that need experimental verification as for reactions above.

14. Repeat model generation and analysis.
15. Analyse the new attractors; repeat QlM update – model generation – model evaluation until the bBM is functional.[57]
16. Iterate until all input-output paths work as expected.
17. In case the QlM model appears to be correct or complete, but the bBM fails to predict the expected input/output behaviour, consider if the bBM export assumptions are appropriate for this particular state or reaction (manually inspect the bBM to determine if there is any reason to redefine that particular update rule).[58]
18. To create a complete file of the simulation results, merge the "[model]_trajectory_first_simulation.csv" files into a single table. Most spread sheet programs can format cells depending on values to generate heat maps such as Figure 5.

**Exempli Gratia: The insulin signalling pathway – part 5.** Our initial model of the insulin pathway turned on but not off. To find out why the signal is interrupted in one direction, we inspect the simulation results in more detail. First, we visualise the simulation results in a heatmap, where states are sorted in order of activation (Fig 5A). We are particularly interested in the difference between steady state 1 (where insulin and outputs are off) and steady state 3 (where insulin is off and output should be off but are on) (Fig 5B). We extract the states and reactions that differ at these two steady states and visualise them in a separate panel (Fig 5C). We can see that in the attractor of round one, all of the phosphorylation states and reactions as well as the reactions depending on those states are OFF. In the attractor of round three these states and reactions are all ON including the output signal [PI3K] and [RAS]. This is interesting, because in round one we simulated the model without the input signal insulin and saw that the output signal does not change, but in round two we simulated with insulin ON and could observe an output response to the input signal, which does not turn OFF again after switching the input signal OFF again. If we have a closer look into the list of states and reactions that are different, we see that all the phosphorylated states and reactions depending on these states are OFF in the first round but ON in the third round. This indicates that the de-phosphorylation reactions required to antagonise the signal are missing. When we added hypothetical de-phosphorylation reactions to our system (table 4), the problem was resolved (Fig 5D & E): The attractor of round one and round three are the same, and the phosphorylated states as well as the downstream states and reactions turn off when insulin is removed. As the model now reproduces – qualitatively – the expected *in vivo* behaviour, we can proceed to the final step: generating the rule based model.

**V. Creation and simulation of the rule-based model**

1. Generate the rule based model.
    a. To create the rule based model from the command line:
       Call the "rxncon2bngl" script. [59]
    b. To create the rule based model through the GUI:
        i. Load the respective project.
        ii. In the top navigation bar choose „Model Export" → „Rule Based Model".
        iii. (Optional) Provide a comment.
        iv. Create the BNGL file by clicking "Create model".
        v. Retrieve the file by clicking the „Show" button in the respective detail view (the eye symbol).
2. This is model generation itself is automated, but the resulting model has trivial parameters (all = 1) and initial amounts (all = 1000).
   To simulate the model run the "BNG2.pl" Perl script with the generated BNGL file as input. By default, it is assumed that the network-free simulator NFsim is used. To use BioNetGen's standard ODE simulation, the statement "simulate_nf" has to be changed to "simulate_ode". For further details we refer to the BioNetGen documentation.
   *Note: For simple systems, ODE simulations will give the best result. However, they require the full network (i.e. all the microstates that can be reached by applying the rules, starting from the initial states) to be generated.*

**Exempli Gratia: The insulin signalling pathway – part 6; creation of the rule based model.** We translated the rxncon model of the insulin pathway into a rule based model in the BioNetGen language**.** The model is available as an electronic supplement (Sup file 5).

*Components in the I-R model*: In the insulin-receptor model there are eight different components. In our NFSim simulations we could easily handle 1000 copies of each component: simulating 150 units of time took only a couple of second on a modest mid-2013 laptop. However, even in this modest system, we have 27 rate constants (excluding the seven dephosphorylation reactions we add later), of which we typically have no precise value.

*Residues and domains*: The elemental states are grouped by component name (i.e. "IR") and within the component name by locus (residue or domain). Take for example the following elemental reaction in the insulin-receptor system: "IR_p+_IR_[TK(Y1158)]", which consumes the state "IR_[TK(Y1158)]-{0}" and produces the state "IR_[TK(Y1158)]-{p}". These two states describe a property of the same residue, and will be translated to different internal states living on a single site of the IR molecule in BNGL. All in all, the Molecule Type declaration for the insulin receptor molecule looks as follows: "IR(IRBDD,JMD,JMDY972R~0~p,TKDY1158R~0~p,TKDY1162R~0~p,TKDY1163R~0~p,ligD)" In rxncon, modifications always happen at the residue level and bindings at the domain level. Therefore, if ambiguity arises due to a name clash, "R" or "D" is appended to the BNGL site name.

*Fully neutral forms*: The fully neutral form, as it appears for the insulin receptor molecule is: "IR(IRBDD,JMD,JMDY972R~0,TKDY1158R~0,TKDY1162R~0,TKDY1163R~0,ligD)" We can identify empty IR, ligand and JM binding domains, and unphosphorylated Y1158, Y1162, Y1163 and Y972 residues.

*The output signals* in the insulin-receptor model are PI3K and the Ras signals. The first is a single bound state between IRS and PI3K, but the second is more interesting. The Ras signal has two contributions: the Shc--Grb2--SOS complex and the IRS--Grb2--SOS complex. These correspond to two different patterns. Since BNGL does not allow algebraic expressions in the *Observables* section, this is solved by appending numbers to the two contributions:

Molecules     PI3K0    IRS(bdD!1).PI3K(SH2D!1)
Molecules     RAS0     Grb2(SH2D!1,SOSD!2).SOS(Grb2D!2).Shc(bd2D!1)
Molecules     RAS1     Grb2(SH2D!1,SOSD!2).IRS(bdD!1).SOS(Grb2D!2)

*Contingencies containing explicit OR-statements or implicit ones* (such as above) require some attention because of the "don't write, don't tell" principle of BNGL. Say that a reaction requires a phosphorylation at site 1 or site 2. If the reaction context of the first rule derived from this reaction has the phosphorylation at site 1, the context of the second rule (where site 2 is phosphorylated) should state explicitly that site 1 is unphosphorylated. We developed a procedure, closely related to the Gram-Schmidt orthogonalisation procedure, to make sure that all rules derived from an elemental reaction have disjunct contexts (in preparation). The rules generated by the rxncon framework automatically all have disjunct reaction contexts.

*Simulation of the rule based model.* Parametrisation and simulation of rule based models fall outside the scope of this chapter, but since we are interested in the response of our system to insulin, we will simulate the following, as explained in detail in the text below:

1. First, we will let the system run for 50 units of time without any insulin present,
2. Then, suddenly, 1000 units of insulin are added to the system. We simulate for another 50 units of time,
3. After that we remove all the insulin and simulate for 50 units of time again.

**Exempli Gratia: The insulin signalling pathway – part 7; simulation of the rule based model.** Studying the Boolean model led to the conclusion that we require extra dephosphorylation reactions to make sure the output signals turn off again when the insulin goes away. However, let us first try to study the response to 1-3 without these dephosphorylation reactions, to see how this works out in the rule-based model.

BioNetGen has a setConcentration command that can adjust species concentrations mid-simulation, however the complete state of such a species has to be specified. For our step 2, the adding of insulin, this is convenient. For step 3, the removal, we will add a degradation rule for insulin controlled by a rate ins_deg_0 whose initial value is 0 and which will be changed to 100 through the setParameter command. The large value of this parameter, compared to the other rate constants, should make sure the insulin disappears almost instantaneously.

The BNG actions at the end of the BNGL file for this setup are:

```
generate_network({overwrite=>1});
simulate_ode({suffix=>"ode_before",t_end=>50,n_steps=>200});
setConcentration("insulin(IRD)", "1000");
simulate_ode({suffix=>"ode_during",t_end=>50,n_steps=>200});
setParameter("ins_deg_0", "100");
simulate_ode({suffix=>"ode_after",t_end=>50,n_steps=>200});
```

After 14 iterations of applying the rules, 2587 species got created and 30839 reactions between them. The network generation took a handful of minutes on a modest laptop. Integrating the ODE system took roughly fifteen minutes.

In Figure 6A, we can see the result of the simulation. Just as in the Boolean system (Fig 5A), the outputs do not switch off after removing the insulin.

**Exempli Gratia: The insulin signalling pathway – part 8; the combinatorial complexity.**

The rule-based simulation plotted in Figure 6 shows the same behaviour as the Boolean simulation in Figure 5: after removal of the insulin, the output signal "stays stuck" in the ON state. We therefore add the same 7 dephosphorylation reactions to the rule-based model. When we try to run the ODE simulation again, this happens:

BioNetGen version 2.2.6
Reading from file insulin_dephos.bngl (level 0)
Read 17 parameters.
Read 15 molecule types.
Read 15 species.
Read 4 observable(s).
Read 49 reaction rule(s).
ACTION: generate_network( insulin_dephos )
Iteration   0:    15 species      0 rxns  0.00e+00 CPU s
Iteration   1:    18 species      4 rxns  2.00e-02 CPU s
Iteration   2:    19 species      8 rxns  0.00e+00 CPU s
Iteration   3:    26 species     17 rxns  2.00e-02 CPU s
Iteration   4:    51 species     75 rxns  7.00e-02 CPU s
Iteration   5:    95 species    273 rxns  2.40e-01 CPU s
Iteration   6:   148 species    665 rxns  5.20e-01 CPU s
Iteration   7:   189 species   1145 rxns  6.10e-01 CPU s
Iteration   8:   290 species   1611 rxns  5.50e-01 CPU s
Iteration   9:   694 species   3019 rxns  1.80e+00 CPU s
Iteration  10:  2889 species  10143 rxns  1.00e+01 CPU s
Iteration  11: 19914 species  61810 rxns  8.56e+01 CPU s
Iteration  12: 125172 species 492015 rxns  9.06e+02 CPU s

After this, we stopped the network generation and reached for NFSim, the network-free simulator. Note how many more species get created by just adding these extra dephosphorylation reactions.

**Exempli Gratia: The insulin signalling pathway – part 9; network free simulation.**

Sadly, the setConcentration command does not work in NFSim. We have found a workaround, which we present here since we could not find it in the NFSim documentation or online. The problem is that we want to, during a simulation, add a number of molecules in a certain state, and at a later time remove them again. Our solution is the following:

First, we add synthesis and degradation rules for the insulin molecule:

| 0 | -> insulin(IRD) | insulin_prod() |
| insulin | -> 0 | insulin_deg() DeleteMolecules |

Here the reaction rates "insulin_prod()" and "insulin_deg()" are function calls we will define below. We furthermore state that the produced insulin will be in the unbound state, and that the degradation rule matches insulin in any state. The "DeleteMolecules" keyword is necessary for our desired behaviour: it says that if we degrade an insulin molecule that is connected into some complex, we should not remove that entire complex from the system, but just the insulin molecule – possibly breaking the complex into subunits along the way.

We furthermore define two parameters:

    ins_prod_0   0
    ins_deg_0    0

and set the seeded concentration of insulin(IRD) to 0.

An observable

  Molecules INSULIN   insulin

which appears in the functions

 insulin_prod() = ins_prod_0 * abs(Numinsulin - INSULIN)
 insulin_deg()  = ins_deg_0  * INSULIN

The INSULIN in these functions is the observable defined above, and Numinsulin is the desired insulin number, set at 1000. We take the absolute value of the difference such that the reaction rate is strictly positive.

Using this functional form for the reaction rates allows us to control the production and degradation of insulin by controlling the values of the parameters "ins_prod_0" and "ins_deg_0", which we can do (see below). By setting ins_prod_0 to a high number (compared to the other rate constants), the production will be active when the number of insulin molecules is smaller than Numinsulin, and when it reaches Numinsulin, the production will be turned off again. By setting ins_deg_0 to a high number the degradation of insulin will be active as long as there is insulin present in the system.

All that is left now is writing a script that performs the simulation and changes the ins_prod_0 and ins_deg_0. For this we first need to generate an XML file that can be read by NFsim by ending the BNGL file with the following action:

  writeXML();

In the current version NFSim's bundled BioNetGen (2.2.2), this crashed since there was a parsing error in the synthesis rule. Using a more recent version of BioNetGen (2.2.6) solved that problem.

The script (a so-called RNF file) contains the following commands, where "insulin.xml" is the just-created XML file.

```
-xml insulin.xml
-v
-o insulin_nf.gdat
begin
  sim 50 200
  set ins_prod_0 100
  update
  sim 50 200
  set ins_prod_0 0
  set ins_deg_0 100
  update
  sim 50 200
end
```

This will simulate the system for 50 units, outputting 200 data points, set the production of insulin to 100, simulate again, set the production of insulin to 0 and the degradation to 100, and simulate again.

The results of the NFSim simulations, both with and without the extra phosphorylation reactions, are given in Figures 6B and C.

# Figure legends

**Figure 1: The network reconstruction process and the levels of information depth.** (A) The conceptual network seed defines the scope and key components. The Input [I] and Output [O] in the network need to be identified, together with the components, here shown as three kinases (K1-K3), that connect the in- and output. Ideally, the pair wise connection between the components can be determined (dashed lines). (B) The reaction layer connects the components. In the next step, the network seed is refined by defining the actual reactions in the network. For this step, the type of reactions between the component pairs (red arrows), as well as the states they produce or consume, need to be identified. (C) The reaction layer lacks causal information. Here, the same information as in (B) is used to display the elemental states that are consumed and produced in each reaction. Note that the network falls apart in isolated motifs. (D) The causal layer connects the reactions and states. The contingencies (red edges) define how reactions (or outputs) depend on states (or inputs). Both the reaction and contingency layers are required to create a connected network at the level of information transfer. The graphs are visualised in SBGN-AF format [21].

**Figure 2: The syntax of elemental reactions and states.** (A) Elemental reactions are defined by one or two components and a reaction type, which are separated by underscores when written as strings. The Components may be specified with domain and/or residues, depending on the reaction type, resulting in states that are defined at the same resolution. The locus (domain and/or residue) are flanked by hard brackets and separated from the component by an underscore. The residue is additionally flanked by normal brackets. For interactions between components (here exemplified by a protein-protein interaction; upper), the resolution becomes elemental (i.e. indivisible and mutually exclusive; see table 1) at the level of domains. For covalent modifications (here exemplified by a phosphorylation; lower), the resolution becomes elemental at the level of residues. The elemental states have the same resolution as reactions. The dimerisation is indicated by a double dash "--" and the phosphorylation by appending "-{P}". Elemental states of one domain or residue are mutually exclusive with other elemental states at the same domain or residue. (B) Contingencies are defined by a reaction (or output) and a combination of elemental states and inputs. Combinations of elemental states can be expressed as Boolean states, which make it possible to define structured complexes. In this example, the binding of insulin requires dimerisation of the receptor but also that insulin has not been bound already (as the receptor dimer only binds one insulin molecule [20]). To express this in rxncon, the reaction IR_[lig]_i_insulin requires the Boolean state <IR-empty>. <IR-Empty> is in turn defined as a complex with two IR monomers at position 0 and 1, where the first is equivalent to the IR in the reaction (the reactants have position 0 and 1 in the namespace of the reaction). The reaction defines that the IR binding insulin must be unbound (a component can only have one bond per

domain), and the contingency defines that it must be in complex with another IR which in turn has no bond at the domain binding insulin. Hence, this defines an IR dimer without insulin bound.

**Figure 3: The insulin model.** (A) Schematic representation of the insulin signalling pathway. Extracellular insulin binds the dimerised insulin receptor (IR), which autophosphorylates and then recruits and phosphorylates the insulin receptor substrate (IRS) and Shc, which in turn, when phosphorylated, binds to Grb2/SOS or the phosphoinositide-3-kinase (PI3K). The figure is adapted from [20]. (B) The rxncon reaction graph visualises the reaction layer of the mechanistic model. Major light green nodes: Components. Minor light green nodes: Domains. Minor dark blue nodes: Residues. Red arrows: Phosphorylations. Grey edges: (Protein-protein) interactions. The network is connected from the top component (insulin) to the two bottom-most components (PI3K and SOS).

**Figure 4: The reaction-contingency model.** The rxncon model visualised as a regulatory graph, which shows elemental reactions (red nodes) and elemental states (blue nodes). The reaction edges (blue arrow; production, purple "tee"-arrow; consumption) define which elemental reactions produce or consume which elemental states. The contingency edges (green arrow; activation, red "tee"-arrow; inhibition (not used in this model) define which elemental states activate or inhibit which elemental reactions. Certain contingencies cannot be defined by individual elemental states; these are defined via Boolean contingencies (white triangles: AND, white diamonds: OR, white octagons: NOT). In this graph, we can follow the path from the most upstream reaction: (IR_[lig]_i_insulin_[IR]through the pathway until it reaches the two outputs: [PI3K] and [Ras]. Hence, the graph appears complete.

**Figure 5: The Boolean simulation results.** (A) The initial model can activate, but not deactivate, the insulin signal. The heatmap shows the state evolution for each elemental reaction and state (rows) over time (columns). The colours indicate the state: Dark grey = true, white = false. The simulation is initiated with the insulin states false, the neutral states of all the other components true and all reactions and the outputs false. The model is simulated until the first steady state is reached (i). (B) As expected, the outputs remain off (B). The neutral state of insulin is set to true, and the simulation repeated until the next steady state is reached (ii). The pathway turns on. To set insulin to false, all insulin states, as well as all reactions that produce them, have to be set to false. With insulin set to false, the simulation is repeated until the next steady state (iii). Surprisingly, the signal does not turn off despite insulin being off, which also means a new steady state not seen before. Hence, the simulation is repeated, this time with insulin on, until the fourth steady state is reached (iv). As expected the outputs are on, and a closer inspection reveals that steady state four is identical to steady state 2 – hence, we have explored the possibilities of the model (at least using synchronous deterministic simulation). (C) To determine why the model failed to shut down, we inspect all the reactions and states that differed between steady state 1 and 3 (grouped in classes to make the list shorter). We see that the first entries in the list are residues that lock in the phosphorylated states.

Closer inspection reveals seven such residues, for which we add dephosphorylation reactions (Table 4). (D) We repeat the simulation with the updated model. The seven new reactions are placed at the bottom. (E) The output now responds as expected to insulin, and steady state 3 is equal to steady state 1 – consequently we only perform three simulations. (F) Closer inspection of the problematic states and reactions from (C) reveals that they all return to the off state in the updated model.

**Figure 6: Simulation of the rule based model.** (A) ODE simulation of the insulin response system in the absence of extra dephosphorylation reactions. (B) NFSim simulation of the insulin response system in the absence of the extra dephosphorylation reactions. The behaviour is similar to the ODE solution in (A), except for the stochastic noise. (C) NFSim simulation of the insulin response system in the presence of the seven extra dephosphorylation reactions. To obtain a signal with a decent signal to noise ratio, the parameters related to the dephosphorylation reactions were chosen as follows: all phosphatases have molecule counts of 10, and the dephosphorylation reactions have a rate constant of 0.025. After insulin switches off, both the AKT and RAS responses switch off. The steady-state response is lower because it requires fully phosphorylated complexes, which are less numerous in the presence of dephosphorylation reactions.

**Table 1: The reaction type definition.** The table defines what constitute well-formed reaction statements, and how these statements translate into a rule in a rule-based model via a skeleton rule. Each entry in these tables provides a definition for a certain type of reactions. The table gives the constraints that the component specifications appearing in a reaction statement have to adhere to in order for the statement to be well-formed. First, the "!MolType" defines which type (i.e. Protein, DNA, mRNA or Any) the reaction operates on. Second, the "!Resolution" specification defines the level of resolution (Component, Domain or Residue). For example, in the definition for a phosphorylation it is stated that the subject has to be a Protein specification at the Component resolution and the object a Protein specification at the Residue resolution. This means that A_p+_B_[(r)] is a valid statement, whereas A_p+_B is not since B is stated at the Component resolution. However, the user can also provide a reaction at a lower resolution than required. In this case, a generic locus name is generated to make the resolution elemental. The "!SkeletonRule" defines the translation of the rxncon statement into a reaction rule, which is subsequently compiled to BNGL. We discuss the details elsewhere [9], but highlight two key features: (1) reactions consume and produce elemental states that live either on molecules or bonds between molecules, (2) we allow for "method calls" on Specification objects that return other Specification objects (i.e. in the translation reaction: $y.to_protein_component_spec returns the ProteinSpecification corresponding to the MRnaSpecification $y). All the "standard" reactions that come out of the box in the rxncon framework are defined precisely in this table, and the user can add further definitions in the ReactionTypeDefinition sheet to be parsed together with the model. Note that new modification types also need to be defined in the ModificationTypeDefinition sheet.

**Table 2: The reaction list of the insulin model.** The reaction layer of the insulin model is defined by 17 elemental reactions, which fall into two classes: complexation (ppi = protein-protein interaction; i = (other) interaction) and covalent modification (P+ = phosphorylation, AP = autophosphorylation). In rxncon, autophosphoryaltion is always within a single molecule, so "autophosphorylation in trans" translate to a normal phosphorylation reaction.

**Table 3: The contingency list of the insulin model.** The contingency layer of the insulin model is defined by 20 contingencies, which spread over 35 lines due to the definition of Boolean states. In this model definition, we only used absolute requirements (!) to define contingencies and the Boolean operators AND and OR to define Boolean states.

**Table 4: The dephosphorylation reactions added to the insulin model in the gap filling process.** The model evaluation showed that the phosphorylated states need to be reversible for the network to be responsive to insulin in both directions. However, the identities of the phosphatase(s) are unknown, hence we add seven unknown phosphatases (uPPase) that may or may not be identical across two or more reactions.

**Acknowledgements**

This work was supported by the German Federal Ministry of Education and Research via e:Bio Cellemental (FKZ0316193, to MK). The chapter has been written for 'Modeling Biomolecular Site Dynamics' (editor William S. Hlavacek), in the series: Methods in Molecular Biology.

# 4. Notes

[1] Anaconda provides an easy way to install the most recent Python, as well as R and RStudio that is used for the bipartite Boolean model analysis. It can be downloaded from: https://www.continuum.io/downloads

[2] The installation process differs slightly between operating systems:

Under Windows:
- Open the console and type "pip install rxncon". The default installation folder will depend on your Python installation. With a python install though Anaconda, the rxncon folder appears in [user]/Anaconda3/lib/Site-packages. The files you will need to call appear in [user]/Anaconda3/Scripts.
- To test the installation, navigate the console to the folder with the scripts and type "python rxncon2bngl.py". Expect a string "Usage: rxncon2bngl.py [OPTIONS] EXCEL_FILE" and an error message "Error: Missing argument "excel_file".

Under OS X:
- Open the console and type "pip install rxncon". The default installation folder will depend on your Python installation. With Anaconda, the rxncon folder appears in [user]/Anaconda3/lib/python3.6/Site-packages. The files you will need to call appear in [user]/Anaconda3/bin.
- To test the installation, navigate the console to the folder with the scripts and type "python rxncon2bngl.py". Expect a string "Usage: rxncon2bngl.py [OPTIONS] EXCEL_FILE" and an error message "Error: Missing argument "excel_file".

Under Linux:
- Make sure you have PIP installed. If not, use your package manager to install it. E.g., on debian-based systems type "sudo apt install python3-pip".
- Open a terminal and type "pip3 install rxncon --user". This installs into $HOME/.local, the executables are in $HOME/.local/bin.
- To get easy access to the rxncon scripts, you can update your PATH environment variable to include this directory: put something like "export PATH=$HOME/.local/bin:$PATH" into your .bashrc.
- To test the installation, type "rxncon2bngl.py". Expect a string "Usage: rxncon2bngl.py [OPTIONS] EXCEL_FILE" and an error message "Error: Missing argument "excel_file".

3. Elemental states are defined by the Specs (Component + Locus) and a state. There are five slightly different ways these states can look:
    i.   Bond state (dimer):          A_[domA]--B_[domB]
    ii.  Bond state (intramolecular): A_[domA]--[domB]
    iii. Unbound domain:              A_[domA]--0
    iv.  Covalent modification:       B_[res]-{mod}
    v.   Unmodified residue:          B_[res]-{0}

    where "A" is componentA's name, "B" is the componentB's name, "domA" and "domB" are the domain names, "res" is the residue name, and "mod" is the modification type.

4. Note that elemental states can be defined without a locus. In this case, it is equivalent to an "OR" statement of all matching elemental states.

5. A bidirectional reaction ("!BidirectionalVerb" = yes) is internally converted into a forward and a reverse reaction. These get unique names by concatenating "!UID:ReactionKey" with "+" and "-". Therefore, these names must also be reserved.

6. An open source R package (v 2.1.1) provided by the cran-r-project (https://cran.r-project.org/; 18.    Mussel, C., M. Hopfensitz, and H.A. Kestler, *BoolNet--an R package for generation, reconstruction and analysis of Boolean networks.* Bioinformatics, 2010. **26**(10): p. 1378-80.).

7. This can also be done through the Anaconda Navigator, if Anaconda is installed, in which case the RStudio install includes R.

8. R can be installed through Anaconda, by opening the console and typing: "conda install –c r r-essentials"

9. In the console, type "R" to enter the R environment. Then type "install.packages("BoolNet")" and select the download server.

10. To install NFSim:
    1. Make sure PERL is installed. *E.g.*, by typing "perl -v" in the console. If not, install PERL first.
    2. Download NFsim from http://michaelsneddon.net/nfsim/download/, and extract the content to a suitable folder. This includes a binary for Windows, Mac and Linux.

3. To test the installation, open the console and navigate to the NFsim folder. Type "perl BNGL.pl –v". Expect "BioNetGen version 2.2".

[11] The scope of the network depends on the objective and aim of your study. Think about which parts of the network are important to include and what is not part of your interest. The scope often changes during the model building, but it is helpful to have a clear idea about the functions and components to be included – even if that is a moving target.

[12] The input and output of a pathway define the border between the detailed mechanistic model and the surrounding cell. Think about the input/output behaviour as the macroscopic function of the pathway, which the detailed molecular mechanisms should explain 19. Hlavacek, W.S. and J.R. Faeder, *The complexity of cell signaling and the need for a new mechanics.* Sci Signal, 2009. **2**(81): p. pe46.. Typical examples of inputs can be signals or cellular states, and examples of outputs include transcriptional activation or cellular decisions. Technically, an input will behave like an elemental state, i.e. it acts as an activator or inhibitor of one or more reaction(s). Correspondingly, the outputs will behave like reactions, i.e. they respond to a (set of) elemental state(s).

[13] This is **not** good practice. The quality of the model generally improves with the number of different sources, and we strongly recommend the use of multiple sources even for this stage and definitely for the mechanistic model building. However, we settle for a single paper as this model is for demonstration purposes only.

[14] The presumed connections are a good starting point to find relevant network information, as they can be used to narrow down literature searches. However, this is only sufficient if the network draft is complete, which is unlikely. It is important to keep in mind that probably other components and reactions between components are needed to connect the model input with the output.

[15] In rxncon, the core reaction definition only includes the component(s) that change and (for some reactions) a catalyst. Complex reactants are defined using contingencies.

[16] New reaction types are defined by adding lines to the ReactionTypeDefinition sheet. The different columns are described briefly in the materials section above, and in detail elsewhere 9. Romers, J.C. and M. Krantz, *rxncon 2.0: a language for executable molecular systems biology.* bioRxiv, 2017.. To extend the list, it is essential to make sure IDs are unique, and that

the molecule type and resolution is consistent with the skeleton rule. In addition, the modification type must be defined in the ModificationTypeDefinition sheet. It is helpful to map new reaction types on the existing ones.

17    This entry must match one of the "!UID:ReactionKey" entries in the ReactionTypeDefinition sheet.

18    The resolution depend on the state(s) that change: Covalent modifications live on residues and bonds live on domains. Catalysts and components that are synthesised or degraded are to be defined at the component level.

19    Intramolecular reactions, such as autophosphorylation or intramolecular bond formation, are defined by entering the same component name in "!ComponentA:Name" and "!ComponentB:Name". Note that the domains and/or residues may differ, though.

20    For catalytic or transport reactions, the "active" component should be specified as Component A. For all reciprocal or non-directional reactions, the order is arbitrary. However, the order must be consistent in all entries: The software will not realise that A bound to B and B bound to A are the same complexes. Hence, we recommend alphabetic order in these cases.

21    Domains are required for bonds. Bonds targeting the same domain in the same molecule are mutually exclusive; hence domain names have an impact on the model generation. If domains are required but not specified, they will be assigned unique names in the parsing step.

22    Domains may be defined when residue information is required. In this case, the residue will live on the domain. This only affects partially undefined contingencies: If the contingency is specified with a phosphorylation at domain resolution only, this will correspond to an or statement phosphorylation on all sites within this domain.

23    Residues are required for covalent modifications. Covalent modifications targeting the same residue in the same molecule are mutually exclusive; hence residue names have an impact on the model generation. If residues are required but not specified, they will be assigned unique names in the parsing step.

24    We prefer using pubmed IDs but any unique identifier would be suitable for references.

| | | |
|---|---|---|
| 25 | | The framework is browser based but a local server is running. All files uploaded to the framework will stay on your local machine, but are saved separately. |
| 26 | | To generate the reaction graph using the console, type: "python [path1]\rxncon2reactiongraph.py [path2]\[model]" where "[path1]" and "[path2]" are the paths to the rxncon scripts and model folder, respectively, and "[model]" is the file name including file extension. |
| 27 | | If you have already laid out an older version of the graph in Cytoscape, you can export this view and use it as a template for upcoming graphs. For this click 'choose file' and select the file with the layout you want before creating the graph. |
| | | If you transfer the layout of an already existing graph (template graph) the graph should be in .xgmml format. The species information in both files will be compared with each other and if there are known coordinates for a certain species in the template graph these will be transferred to the new graph. All species which are not mentioned in the template graph will be visualised centred in Cytoscape. |
| 28 | | The rxncon2cytoscale.xlm style file can be downloaded from: https://github.com/rxncon/tools. |
| 29 | | There are automatic layouts provided by Cytoscape. Click on Layout in the upper Menu -> choose a layout -> All Nodes. This will give you a good starting point you can further proceed from. |
| 30 | | The elemental reactions and states (that are defined in the reaction layer) are needed to define the contingencies in the causal layer. Hence, the reaction layer is a prerequisite for the causal layer (see figure 1). However, in practice, the model building goes back and forth between reactions and contingencies. |
| 31 | | Contextual information is not always easy to find, as it takes dedicated experiments to identify all the requirements. Depending on the number of modification sites and interaction partners, these experiments can be time and resource demanding. Consequently, we expect many important contingencies to remain unknown. |

| | |
|---|---|
| 32 | This can be the existence of the component itself. However, components are typically regulated in information transfer networks, meaning that the state of the component determines its activity. |
| 33 | Keep in mind that for a functional pathway, the active forms of the components need to be reversed when the signal is not present anymore. For this, the reverse reactions will be needed to reset the components. These are implicit for interaction reactions, but not for covalent modification or synthesis/degradation reactions. |
| 34 | There are six contingency symbols. Two absolute: ! = required; x = inhibitory. Two quantitative K+ = stimulation, K- = inhibition. Two with no effect: 0 = known to have no effect, ? = no known effect. The last two are equivalent for the interpretation of the model. |
| 35 | Boolean contingencies can be defined in the same format for convenience. If so, the target is a Boolean state (defined by name within pointy brackets < >), the modifier elemental states, Boolean states, or inputs; and the contingency symbol is the type of Booelan: AND, OR, NOT. Note that a single Boolean state can only take one Boolean expression: Use nested Booleans to create mixed statements. |
| 36 | Input and outputs defined the borders between the mechanistic rxncon model and the surrounding. Note that they can also be used as borders to internal processes that are not well known. By assigning the same name to an input and an output, these are made equivalent, essentially forming a feedback into the pathway. The effect of this varies with modelling formalism. |
| 37 | The components listed in the reaction always have indices 0 (component A) and 1 (component B). It is imperative that all indices in the contingencies are consistent with these, and with each other. Note that each Boolean has its own namespace, and equivalences must be declared when a Boolean is used. For examples, see the contingency list in the insulin example (Table 3). |
| 38 | Equivalences must be defined in structured complexes or when the component name is an ambiguous identifier (e.g. in homodimers or reactions between two copies of the same component). For examples, see the contingency list of the insulin example (Table 3). |
| 39 | The graph can be created from the command line: "python [path1]\rxncon2regulatorygraph.py [path2]\[model]" from the console. In the GUI, the "regulatory graph" button should be selected. For all other purposes, see the reaction graph creation. |
| 40 | The regulatory graph is a simplified view of the regulatory structure of the rxncon model. Alternatively, the full set of states and reactions can be visualised with the species-reaction graph. The graph can be created from the command line: "python [path1]\rxncon2srgraph.py [path2]\[model]" from the console. In the GUI, the "species-reaction graph" button should be selected. For all other purposes, see the reaction graph creation. |
| 41 | The visualisation process is the same as for the reaction graph, except that the style chosen should be rxncon_regulatorygraph. |
| 42 | The QlM you created can be evaluated using a Boolean model. In this modelling approach, you assume that each node in your model can have only two states: Either True/1/on or False/0/off. Each node is defined by a specific update function, describing the dependency on other nodes of the system. At each time step $t$ the rule for a certain component is evaluated by substituting the components within a rule by their states. The result will be the state of this component at $t+1$. Here, we create update rules for state and reactions, rather than components (the typical approach). This has two nice advantages: First, we can derive a unique model (with defined truth tables) from each rxncon network. Second, we can distinguish different active states. |
| 43 | Type: "python [path1]\rxncon2boolnet.py [path2]\[model]" where "[path1]" and "[path2]" are the paths to the rxncon scripts and model folder, respectively, and "[model]" is the file name including file extension. |
| 44 | The different options for the Boolean model generation are listed by calling the program with the "--help" option. They can be selected by calling the command line program with certain flags. Currently, options exit to:<br>• control smoothing the availability of source states in time,<br>• express knockouts or overexpressions,<br>• choose whether to interpret positive influences / negative influences as strict requirements / strict inhibitions or to ignore them, and<br>• select an output filename. |
| 45 | The same options as above are available. |

46   Typically, these models are too large for an exhaustive scan of initial configurations, as the number of possible starting states is $2^n$, where n equals the number of nodes. Instead, we start from a generic start position where all components without states are on, all neutral states are on, and all other nodes are off. See 12.    Thieme, S., et al., *Bipartite Boolean modelling - a method for mechanistic simulation and validation of large-scale signal transduction networks.* bioRxiv, 2017. for details. From this starting point, we typically perform a simulation to let the system find its own natural initial state, before starting to change the input configuration.

47   Here, we show how to simulate and plot a network in R-studio (https://www.rstudio.com/products/rstudio/download2/) using the BoolNetSim.R script that can be downloaded from GitHub (https://github.com/rxncon/tools). The R-script contains some comments for your convenience. You can also run the commands line by line.  Change the R-script to adapt it to your needs.

48   Alternatively, define the path to the model files when setting the model prefix (below).

49   Where [model] is the file name of the rxncon model without file extension. This should also include a path if your file is located in a different directory than the R-script.

On some machines, the script may not open correctly. If so, simply copy/paste the text from the script into R-Studio, and define the file prefix with path to the model file.

50   The script executes a second simulation from the first attractor to test whether it is a point or cyclic attractor. In the first case, the second simulation results consist of two identical columns (starting point, and final attractor). In the second case, the simulation results consist of more than two columns that change.

51   This is either the point attractor or a single state in a cyclic attractor, depending on the model.

52   The key to the symbols can be found in the "[model]_symbols" file.

53   As we use deterministic updates, we have seen all states we can reach from these input configurations.

54. It is important to check this in both directions. It is easy to create a model that only responds in one direction, or only once. Therefore, the input needs to be varied until the model returns to a previous steady state – and the output should still be responsive. As can be seen in the insulin example, this is not necessarily the case even when the regulatory graph appears connected and the output responds the first time.

55. The heatmap and the regulatory graph are powerful tools to help this search. The (sorted) heatmap can be used to track down the first reactions / states that do not change as expected, and the regulatory graph summarise the complete regulatory information at a graphical level.

57. If one type of error is detected (e.g. missing phosphorylation), they are likely to appear elsewhere too. Hence, it makes sense to visually inspect the regulatory graph for other possible sources of similar problems before repeating the complete cycle. In one case, we needed to add fifty hypothetical dephosphorylation reactions (15.Flottmann, M., et al., *Reaction-contingency based bipartite Boolean modelling.* BMC Syst Biol, 2013. **7**: p. 58.).

58. In some cases, the default assumptions in the bBM generation may be inappropriate. If so, this could prevent the model from passing the validation step. If this appears to be the case, adapt the bBM manual to the appropriate update rules.

59. To call the script from the command line, type "python [path1]\rxncon2bngl.py [path2]\[model]" with the same paths as for Boolean model export above.

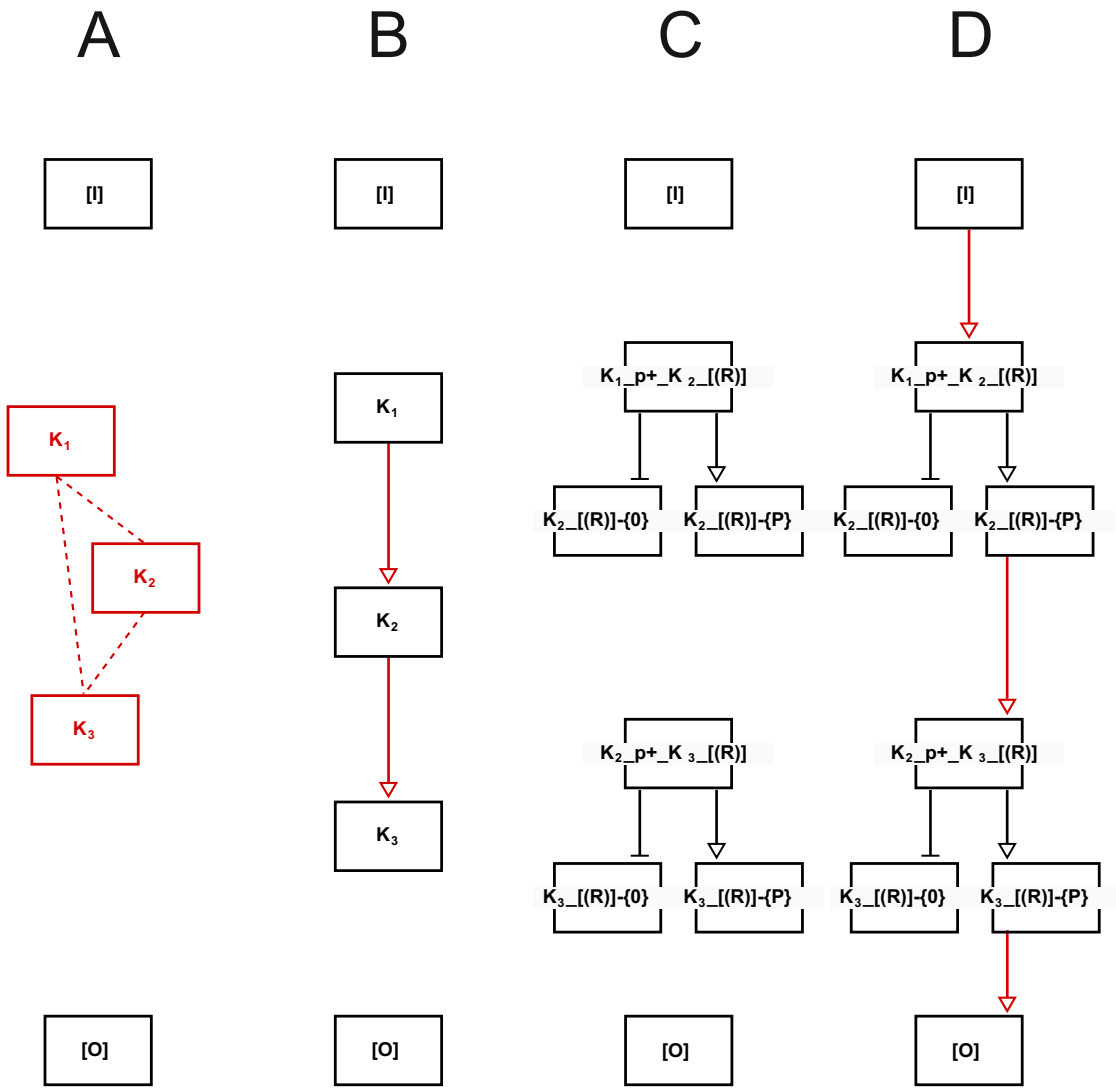

Figure 1

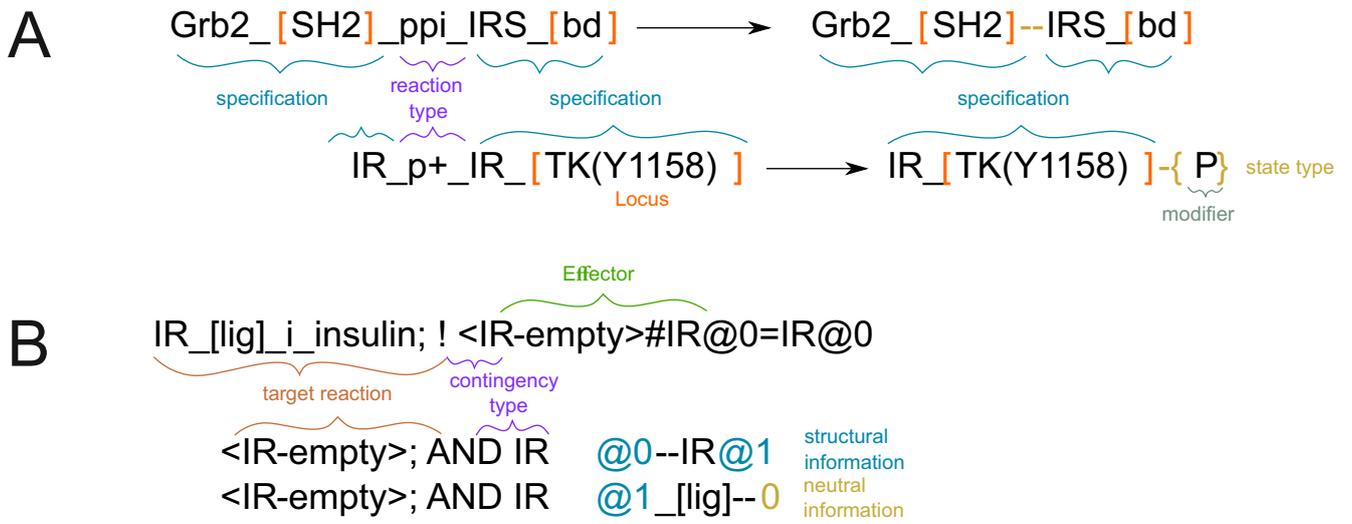

Figure 2

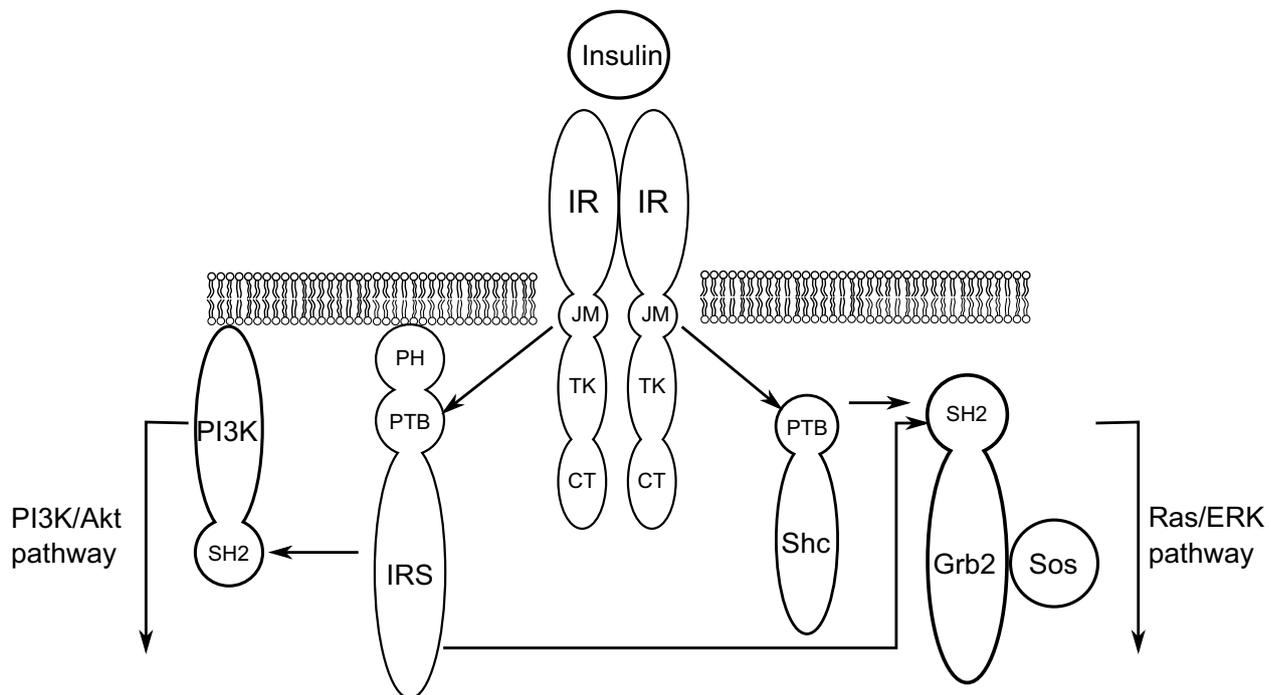
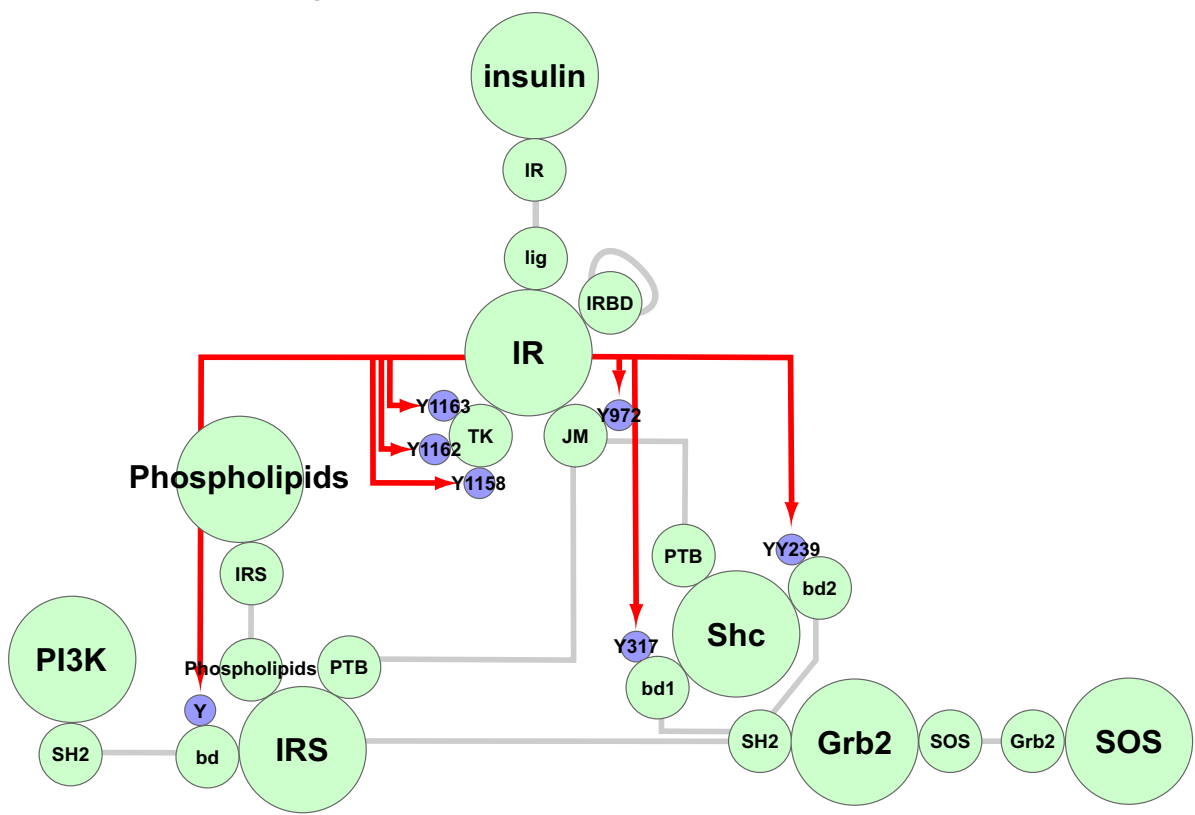

Figure 3

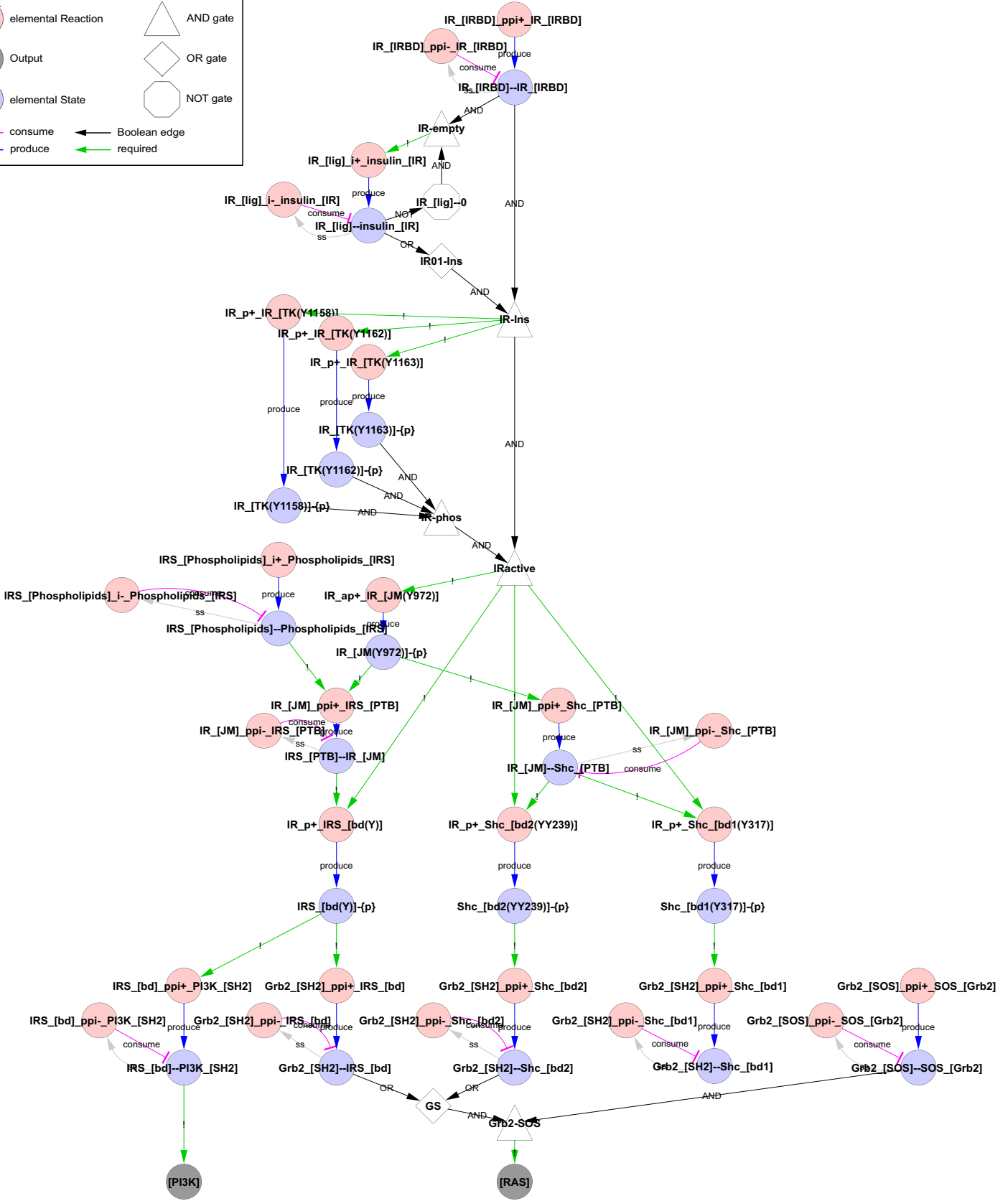

Figure 4

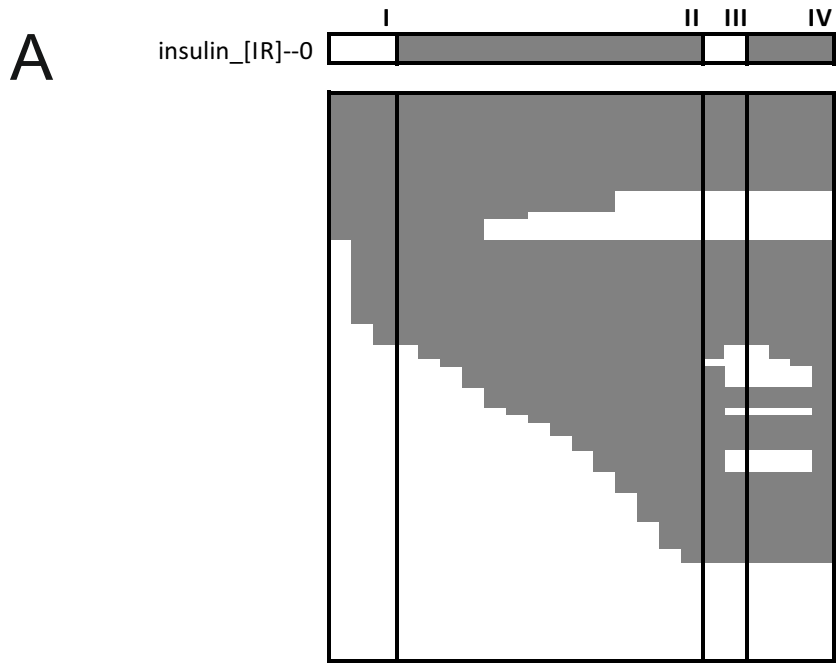
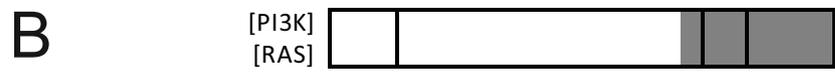
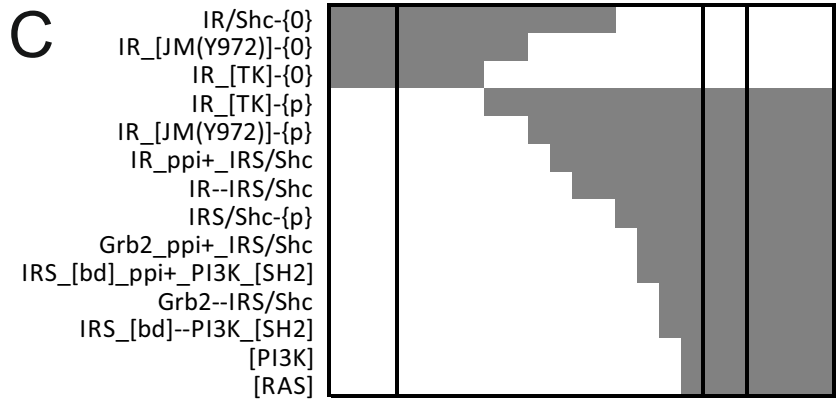
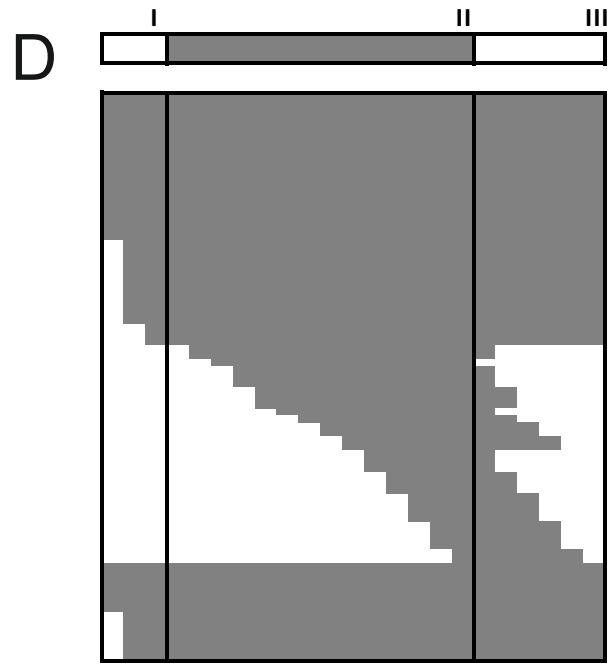
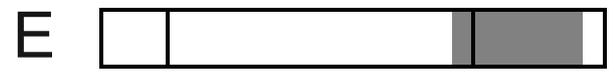
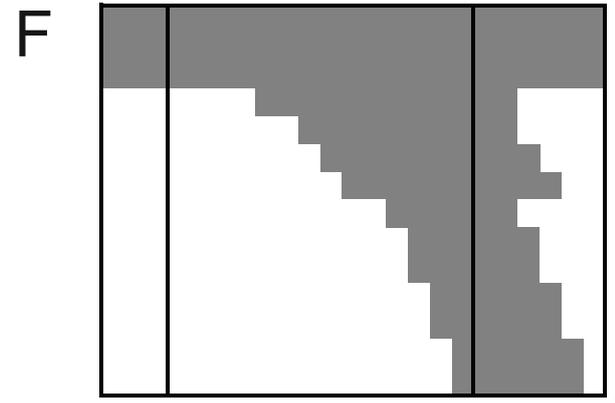

Figure 5

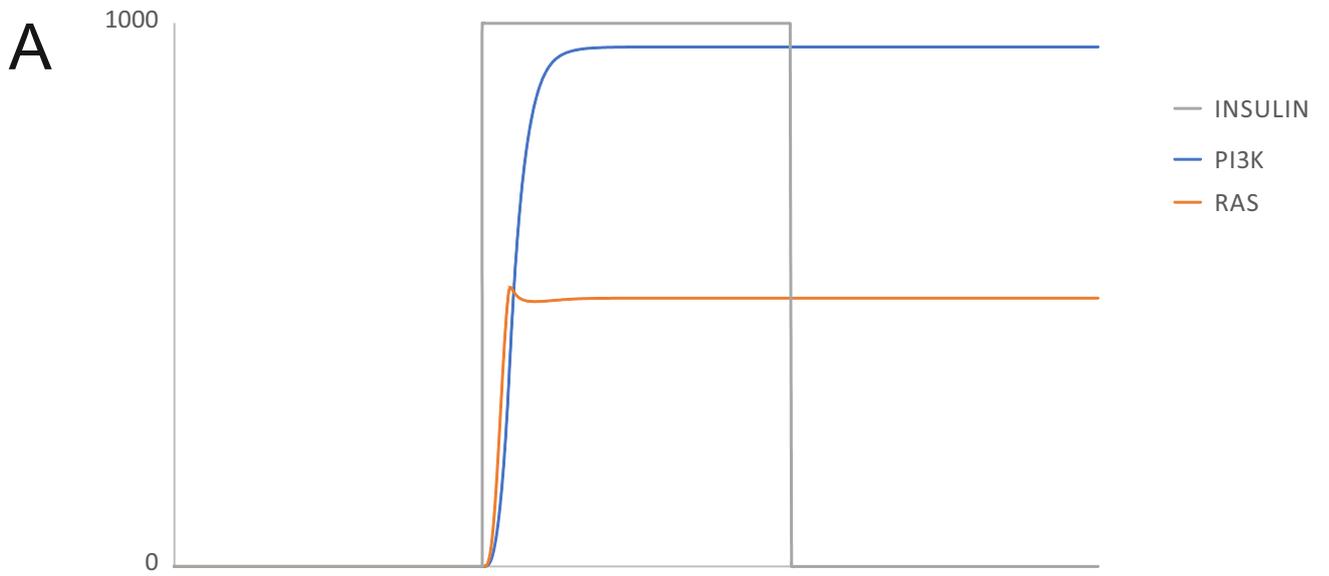
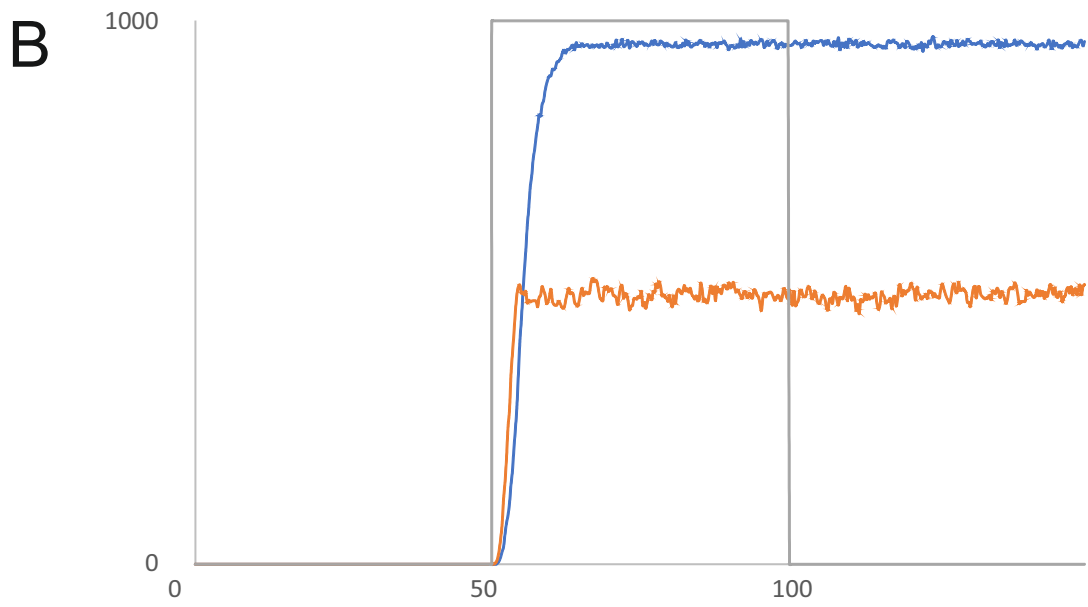
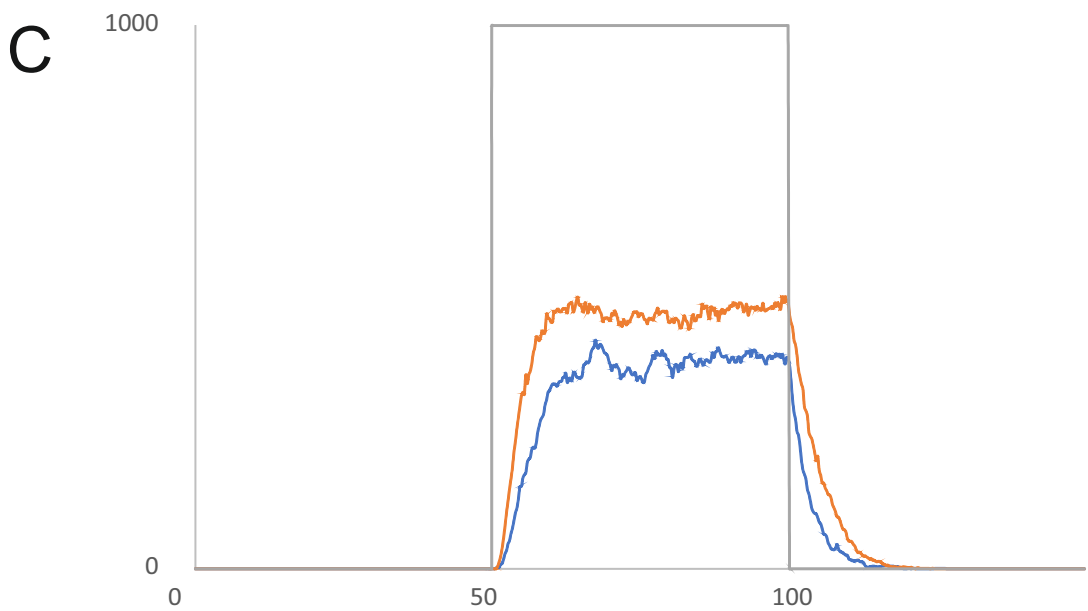

Figure 6

| !UID:Reaction | !UID:ReactionKey | !BidirectionalVerb | !MolTypeX | !ResolutionX | !MolTypeY | !ResolutionY | !SkeletonRule |
|---|---|---|---|---|---|---|---|
| phosphorylation | p+ | no | Protein | component | Protein | residue | $x%#  + $y%#$y%-{0} -> $x%# + $y%#$y%-{p} |
| dephosphorylation | p- | no | Protein | component | Protein | residue | $x%# + $y%#$y%-{p} -> $x%# + $y%#$y%-{0} |
| auto-phosphorylation | ap+ | no | Protein | component | Protein | residue | $y%#$y%-{0} -> $y%#$y%-{p} |
| phosphotransfer | pt | no | Protein | residue | Protein | residue | $x%#$x%-{p} + $y%#$y%-{0} -> $x%#$x%-{0} + $y%#$y%-{p} |
| guanine-nucleotide-exchange | gef | no | Protein | component | Protein | residue | $x%# + $y%#$y%-{0} -> $x%# + $y%#$y%-{GTP} |
| GTPase-activation | gap | no | Protein | component | Protein | residue | $x%# + $y%#$y%-{GTP} -> $x%# + $y%#$y%-{0} |
| ubiquitination | ub+ | no | Protein | component | Protein | residue | $x%# + $y%#$y%-{0} -> $x%# + $y%#$y%-{ub} |
| truncation | cut | no | Protein | component | Protein | residue | $x%# + $y%#$y%-{0} -> $x%# + $y%#$y%-{truncated} |
| protein-protein-interaction | ppi | yes | Protein | domain | Protein | domain | $x%#$x%--0 + $y%#$y%--0 -> $x%!$y%#$x%--$y% |
| intra-protein-interaction | ipi | yes | Protein | domain | Protein | domain | $x%#$x%--0!$y%--0 -> $x%#$x%--[$y.locus%] |
| interaction | i | yes | Any | domain | Any | domain | $x%#$x%--0 + $y%#$y%--0 -> $x%!$y%#$x%--$y% |
| protein-gene-interaction | bind | yes | Protein | domain | Gene | domain | $x%#$x%--0 + $y%#$y%--0 -> $x%!$y%#$x%--$y% |
| transcription | trsc | no | Protein | component | Gene | component | $x%# + $y%# -> $x%# + $y%# + $y.to_mrna_component_spec()%#0 |
| translation | trsl | no | Protein | component | mRNA | component | $x%# + $y%# -> $x%# + $y%# + $y.to_protein_component_spec()%#0 |
| synthesis | syn | no | Protein | component | Any | component | $x%# -> $x%# + $y%#0 |
| degradation | deg | no | Protein | component | Any | component | $x%# + $y%# -> $x%# |

Table 1

| !ComponentA: Name | !ComponentA: Domain | !ComponentA: Residue | !Reaction | !ComponentB: Name | !ComponentB: Domain | !ComponentB: Residue |
|---|---|---|---|---|---|---|
| IR | IRBD | | ppi | IR | IRBD | |
| IR | lig | | i | insulin | | |
| IR | | | p+ | IR | TK | Y1158 |
| IR | | | p+ | IR | TK | Y1162 |
| IR | | | p+ | IR | TK | Y1163 |
| IR | | | ap+ | IR | JM | Y972 |
| IR | JM | | ppi | IRS | PTB | |
| IRS | | | i | Phospholipids | | |
| IR | | | p+ | IRS | bd | Y |
| Grb2 | | | ppi | SOS | | |
| Grb2 | SH2 | | ppi | IRS | bd | |
| IR | JM | | ppi | Shc | PTB | |
| IR | | | p+ | Shc | bd2 | YY239 |
| IR | | | p+ | Shc | bd1 | Y317 |
| Grb2 | SH2 | | ppi | Shc | bd2 | |
| Grb2 | SH2 | | ppi | Shc | bd1 | |
| IRS | bd | | ppi | PI3K | SH2 | |

Table 2

| !Target | !Contingency | !Modifier |
|---|---|---|
| IR_[lig]_i_insulin | ! | <IR-empty>#IR@0=IR@0 |
| <IR-empty> | AND | IR@0--IR@1 |
| <IR-empty> | AND | IR@1_[lig]--0 |
| IR_p+_IR_[TK(Y1158)] | ! | <IR-Ins>#IR@0=IR@0#IR@1=IR@1 |
| <IR-Ins> | AND | IR@0--IR@1 |
| <IR-Ins> | AND | <IR01-Ins>#IR@0=IR@0#IR@1=IR@1 |
| <IR01-Ins> | OR | IR@0_[lig]--insulin@2 |
| <IR01-Ins> | OR | IR@1_[lig]--insulin@3 |
| IR_p+_IR_[TK(Y1162)] | ! | <IR-Ins>#IR@0=IR@0#IR@1=IR@1 |
| IR_p+_IR_[TK(Y1163)] | ! | <IR-Ins>#IR@0=IR@0#IR@1=IR@1 |
| IR_ap+_IR_[JM(Y972)] | ! | <IRactive>#IR@0=IR@0 |
| <IRactive> | AND | <IR-phos>#IR@0=IR@0 |
| <IRactive> | AND | <IR-Ins>#IR@0=IR@0 |
| <IR-phos> | AND | IR@0_[TK(Y1158)]-{P} |
| <IR-phos> | AND | IR@0_[TK(Y1162)]-{P} |
| <IR-phos> | AND | IR@0_[TK(Y1163)]-{P} |
| IR_[JM]_ppi_IRS_[PTB] | ! | IR_[JM(Y972)]-{P} |
| IR_[JM]_ppi_IRS_[PTB] | ! | IRS--Phospholipids |
| IR_p+_IRS_[bd(Y)] | ! | IR_[JM]--IRS_[PTB] |
| IR_p+_IRS_[bd(Y)] | ! | <IRactive>#IR@0=IR@0 |
| Grb2_[SH2]_ppi_IRS_[bd] | ! | IRS_[bd(Y)]-{P} |
| IR_[JM]_ppi_Shc_[PTB] | ! | IR_[JM(Y972)]-{P} |
| IR_p+_Shc_[bd2(YY239)] | ! | IR_[JM]--Shc_[PTB] |
| IR_p+_Shc_[bd2(YY239)] | ! | <IRactive>#IR@0=IR@0 |
| IR_p+_Shc_[bd1(Y317)] | ! | IR_[JM]--Shc_[PTB] |
| IR_p+_Shc_[bd1(Y317)] | ! | <IRactive>#IR@0=IR@0 |
| Grb2_[SH2]_ppi_Shc_[bd2] | ! | Shc_[bd2(YY239)]-{P} |
| Grb2_[SH2]_ppi_Shc_[bd1] | ! | Shc_[bd1(Y317)]-{P} |
| IRS_[bd]_ppi_PI3K_[SH2] | ! | IRS_[bd(Y)]-{P} |
| <Grb2-SOS> | AND | Grb2--SOS |
| <Grb2-SOS> | AND | <GS> |
| <GS> | OR | Grb2_[SH2]--Shc_[bd2] |
| <GS> | OR | Grb2_[SH2]--IRS_[bd] |
| [RAS] | ! | <Grb2-SOS> |
| [PI3K] | ! | IRS_[bd]--PI3K_[SH2] |

Table 3

| !ComponentA: Name | !ComponentA: Domain | !ComponentA: Residue | !Reaction | !ComponentB: Name | !ComponentB: Domain | !ComponentB: Residue |
|---|---|---|---|---|---|---|
| uPPase1 | | | p- | IR | TK | Y1158 |
| uPPase2 | | | p- | IR | TK | Y1162 |
| uPPase3 | | | p- | IR | TK | Y1163 |
| uPPase4 | | | p- | IR | JM | Y972 |
| uPPase5 | | | p- | IRS | bd | Y |
| uPPase6 | | | p- | Shc | bd2 | YY239 |
| uPPase7 | | | p- | Shc | bd1 | Y317 |

Table 4